\documentclass[10pt,a4paper,superscriptaddress,aps,prd,showkeys,showpacs,nofootinbib,reprint,preprintnumbers]{revtex4-2}
\usepackage[utf8]{inputenc}
\usepackage{verbatim}
\usepackage{amsmath,amsfonts,amssymb}
\usepackage[breaklinks,colorlinks]{hyperref}
\usepackage{natbib}
\usepackage{tikz}
\usepackage{float}
\usepackage{graphicx}
\usepackage{adjustbox}
\usepackage{xcolor}
\usepackage{ulem}
\usepackage{cancel}

\usepackage{tabularx}
\usepackage{amsbsy}
\usepackage{braket}
\hypersetup{%
,urlcolor=blue
,citecolor=blue
,linkcolor=blue
}

\def\aap{Astronomy and Astrophysics}
\def\apjl{The Astrophysical Journal Letters}
\def\physrep{Physics Reports}

\def\mnras{Monthly Notices of the Royal Astronomical Society}

\def\aplett{Astrophysical Letters}

\def\jcap{Journal of Cosmology and Astro-Particle Physics}
\def\pasa{Publications of the Astronomical Society of Australia}
\def\apjs{The Astrophysical Journal Supplementary Series}

\def\half{\textstyle{1\over 2}}

\begin{document}

\title{Searching for Time-Dependent Axion Dark Matter Signals in Pulsars}

\author{R.~A.~Battye}
\email[]{richard.battye@manchester.ac.uk}
\affiliation{%
Jodrell Bank Centre for Astrophysics, School of Natural Sciences, Department of Physics and Astronomy, University of Manchester, Manchester, M13 9PL, U.K.
}

\author{M.~J.~Keith}
\email[]{michael.keith@manchester.ac.uk}
\affiliation{%
Jodrell Bank Centre for Astrophysics, School of Natural Sciences, Department of Physics and Astronomy, University of Manchester, Manchester, M13 9PL, U.K.
}

\author{J.~I.~McDonald}
\email[]{jamie.mcdonald@uclouvain.be}
\affiliation{Centre for Cosmology, Particle Physics and Phenomenology,
Université Catholique de Louvain,
Chemin du cyclotron 2,
Louvain-la-Neuve B-1348, Belgium}
\preprint{IRMP-CP3-23-14}

\author{S.~Srinivasan }
\email[]{sankarshana.srinivasan@manchester.ac.uk}
\affiliation{%
Jodrell Bank Centre for Astrophysics, School of Natural Sciences, Department of Physics and Astronomy, University of Manchester, Manchester, M13 9PL, U.K.
}

\author{B.~W.~Stappers}
\email[]{ben.stappers@manchester.ac.uk}
\affiliation{%
Jodrell Bank Centre for Astrophysics, School of Natural Sciences, Department of Physics and Astronomy, University of Manchester, Manchester, M13 9PL, U.K.
}
\author{P.~Weltevrede}
\email[]{patrick.weltevrede@manchester.ac.uk}
\affiliation{%
Jodrell Bank Centre for Astrophysics, School of Natural Sciences, Department of Physics and Astronomy, University of Manchester, Manchester, M13 9PL, U.K.
}

\label{firstpage}

\date{\today}

\begin{abstract}
Axion dark matter can be converted into photons in the magnetospheres of neutron stars leading to a spectral line centred on the Compton wavelength of the axion. Due to the rotation of the star and the plasma effects in the magnetosphere the signal is predicted to be periodic with significant time variation that persists across phase but is narrow in frequency - a unique smoking gun for axion dark matter. As a proof of principle and to develop the methodology, we carry out the first time domain search of the signal using data from PSR J2144$-$3933 taken as part of the MeerTIME project on MeerKAT telescope. We search for specific signal templates using a matched filter technique and discuss when a time-domain analysis (as is typically the case in pulsar observations) gives greater sensitivity to the axion-coupling to photons in comparison to a simple time-averaged total flux study. We do not find any candidate signals and, hence,  impose an upper limit on the axion-to-photon coupling of $g_{a\gamma\gamma}<5.5\times 10^{-11}(D/0.165\,{\rm pc})\,{\rm GeV}^{-1}$ where $D$ is the pulsar distance, over the mass range $m_{\rm a}=3.9-4.7\,\mu{\rm eV}$ using this data. This limit relies on PSR J2144$-$3933 not being an extremely aligned rotator, as strongly supported by simple arguments based on the observed pulse profile width. We discuss the possibilities of improving this limit using future observations with MeerKAT and also SKA1-mid and the possibility of using other objects.  
Finally, to evade modelling uncertainties in axion radio signals, we also carry out a generic ``any periodic-signal search" in the data, finding no evidence for an axion signal. 
\end{abstract}



\maketitle

\section{Introduction}

The search for dark matter in the form of axions~\cite{ref:PQ,ref:K,ref:SVZ,ref:DFSZ,ref:Zhit,ref:misalign1,ref:misalign2,ref:misalign3,Arvanitaki:2009fg,Svrcek:2006yi} is continuing to gather momentum at a dramatic pace. Of particular interest is the mass range $m_{\rm a}\approx 0.1-1000\,\mu{\rm eV}$ where plausible scenarios~\cite{Davis:1989nj,Battye:1994au,  ref:Battye,ref:WS,Bae:2008ue,ref:Marsh,Borsanyi:2016ksw,Choi:2020rgn} have been proposed to realise the Cold Dark Matter abundance $\Omega_{\rm c}h^2\approx 0.12$ that is found to be compatible with cosmological observations, for example, those of the CMB~\cite{ref:Planck2018}. There are a number of haloscope experiments \cite{sikivie1983} which have placed constraints on the axion coupling to photons. These include early cavity experiments \cite{DePanfilis:1987dk,Wuensch:1989sa,Hagmann:1990tj} and their modern incarnations including ADMX \cite{ADMX:2009iij,ref:ADMX2018,ADMX:2009iij,ADMX:2018gho,ADMX:2019uok,ADMX:2021nhd} and its various upgrades and pathfinders \cite{ADMX:2018ogs,ADMX:2021mio,Crisosto:2019fcj}. This has spawned a plethora of active and proposed experiments including CAPP \cite{CAPP:Lee_2020cfj,CAPP:Jeong_2020cwz,CAPP:Lee_2022mnc,CAPP:Kim2022}, HAYSTAC \cite{HAYSTAC:Brubaker2017, HAYSTAC:2018rwy,  HAYSTAC:2020kwv} QUAX \cite{Alesini:2019ajt, Alesini:2020vny}, ORGAN \cite{McAllister:2017lkb,Quiskamp:2022pks}, CAST-RADES \cite{CAST:2020rlf}, TASEH \cite{TASEH:2022vvu} and GrAHal \cite{Grenet:2021vbb} aiming to detect axions using similar techniques. Complementing this, there are also a number of proposed experiments which go beyond the cavity paradigm, allowing laboratory searches to achieve broad frequency coverage across previously challenging frequency ranges. These include
plasma haloscope designs like ALPHA \cite{Millar:2022peq}, broadband reflectors envisaged in the BREAD \cite{BREAD:2021tpx} collaboration, and dielectric haloscopes such as MADMAX \cite{Beurthey:2020yuq}. There are also novel designs which aim to detect magnetic fields induced by axion sources such as DMRadio \cite{DMRadio:2022pkf},  which will operate at lower frequencies ($m_{\rm a}\lesssim \mu {\rm eV}$) and those seeking to using novel materials at higher frequencies \cite{Marsh:2022fmo,McDonald:2021hus,Schutte-Engel:2021bqm}

The largest magnetic fields currently used in laboratory searches for axion dark matter typically do not exceed $\sim 10^5 {\rm G } \, (10\,{\rm T})$, a limiting factor in such searches. By contrast, astrophysical magnetic fields in neutron stars can be as high as $\sim 10^{15} {\rm G } \, ( 10^{11}\,{\rm T})$, making them excellent targets for indirect searches of axion dark matter \cite{Pshirkov:2007st,Huang:2018lxq,hook2018}. In addition, neutron stars are surrounded by a magnetosphere whose varying plasma frequency matches the axion mass across a broad range of masses. This degeneracy leads to a dramatic resonant enhancement of the signal emanating from regions with $m_a \simeq \omega_{\rm p}$, where $\omega_{\rm p}$ is the plasma frequency. As a result, neutron stars can act as broadband axion dark matter detectors. 

Based on a simple but representative model for a neutron star magnetosphere and the density of axions around the star, Refs.~\cite{pshirkov2009,hook2018,Huang:2018lxq} predicted signals that could be easily detected using current and future telescopes operating in the radio-mm waveband, which corresponds to the Compton wavelength of the dark matter axion scenarios referred to above. Spurred on by this, great progress has recently been made in characterising the signal properties using sophisticated ray-tracing methods~\cite{Leroy:2019ghm,Battye:2021xvt,Witte:2021arp} which are capable of computing the line width induced from plasma effects and the precise time-variation and angular dependence of the signal. Early attempts have also been made to address axion-photon mixing in 3D~\cite{Battye_2020,Millar2021}, though this remains an ongoing area of research. 

Various searches have been carried out to detect radio signals produced by axion dark matter converting into photons in the magnetospheres of neutron stars using the Goldreich-Julian (GJ) model~\cite{goldreich1969} for the magnetosphere and estimates of the local density of dark matter, extrapolated to the location of the star in question. These searches have either looked for a background excess near the Galactic centre from populations of neutron stars~\cite{Safdi2019,FosterSETI2022}, or have focused on single objects such as the Galactic centre magnetar~\cite{Darling:2020plz,Darling:2020uyo,Battye2022} or isolated neutron stars~\cite{Foster:2020pgt,Zhou:2022yxp}, and have established bounds on the axion-to-photon coupling, $g_{\rm a\gamma\gamma}$, which are better than the bounds from the axion helioscope CAST~\cite{ref:CAST}. 

Now that the first wave of searches has been carried out, a natural question to ask is what improved observational strategies are available to increase our sensitivity to the axion photon coupling and boost our chances of detecting dark matter axions from neutron stars. In this vein, one might also ask how our newly attained understanding of precise signal properties (including time and frequency information) might be leveraged to increase the power of such searches. To date, all the searches for axions using neutron stars have focused on looking for a spectral line in the frequency domain. The goal of the present work is twofold: (i) to establish a framework of time-domain searches for axion dark matter signals in radio data and (ii) to demonstrate this technique by carrying out time-domain observations of pulsars. Our goal is to understand under what circumstances augmenting these searches to include time-domain information of the signal can improve sensitivity to $g_{a \gamma \gamma}$ and by injecting signal templates into radio data, demonstrate a practical route to obtaining  limits on $g_{a \gamma \gamma}$ which out-perform a simple line-search in the frequency domain. 

The structure of the paper is organised as follows. In sec \ref{sec:SignalProperties} we review the mechanism for photon production from axion dark matter, describe our ray-tracing procedure for modelling the radio signal and discuss the time-dependence of the expected signal. In sec.~\ref{sec:timeDomain} we describe a procedure to search for time-dependent signals in data using a matched filter, and use this to outline what types of periodic signals lead to a strong gain from time-domain information.  In sec.~\ref{sec:Observations} we apply our pipeline to MeerKAT observations of PSR J2144$-$3933 to search for axion dark matter. Our null result is used to place limits on the axion coupling to photons. In sections \ref{sec:future} and \ref{sec:periodic} we explore possible future targets and perform a generalized search for periodic signals. In section \ref{sec:conclusions} we offer our conclusions. 

\section{Modelling the signal due to axions}\label{sec:SignalProperties}

The conversion between axions and photons in strong magnetic fields was laid out in the classic reference \cite{Raffelt:1987im}. It was pointed out in \cite{Pshirkov:2007st} that this mixing could convert dark matter axions into radio photons\footnote{See \cite{Hardy:2022ufh} for a similar mechanism with dark photons.} in the strongly magnetised plasmas which surround neutron stars. In recent years, as axions have moved to the forefront as dark matter candidates, these ideas have been pursued with renewed vigour \cite{hook2018,ref:NS-Japan} and this programme has lead to a variety of observations \cite{darling2020prl,darling2020apj,Battye2022,foster2020,FosterSETI2022} searching for radio lines from dark matter axions. These observational efforts have been accompanied by a more concerted effort to improve the modelling of the signal itself \cite{leroy2020,Witte:2021arp,Battye:2021xvt,Millar2021,Carenza:2023nck} which consists primarily of developing ray-tracing packages to precisely track the photons from their point of emission to the observer, thereby allowing one to derive signal templates which could be detected by a radio telescope.  We now briefly review the basic features of the production mechanism and ray-tracing routine. More details can be found in \cite{leroy2020,Witte:2021arp,Battye:2021xvt,McDonaldInPrep}.

\subsection{Ray-tracing photons in magnetised plasmas}

Resonant conversion between axions and photons occurs at points where $k^\mu_\gamma = k_a^\mu$ where $k^\mu_\gamma$ and $k_a^\mu$ are the photon and axion 4-momentum, respectively. An attempt was made to understand the conversion probability for axions to photons $p_{a \gamma}$ in \cite{Millar2021} leading to 
\begin{equation}\label{eq:AxiontoPhoton}
    p_{a \gamma \gamma} = \frac{\pi}{2} \frac{g_{a\gamma \gamma}^2 \sin^2 \theta_B \left| \textbf{B} \right|^2}{ \left|\textbf{k}_\gamma \right| \left| \omega_p^\prime \right|} 
    \cdot \frac{ m_a^5}{\left(\left|\textbf{k}_\gamma \right|^2 + m_a^2 \sin^2 \theta_B \right)^2},
\end{equation}
which attempts to incorporate 3D effects into the conversion probability. This characterises the ratio of the energy density between an axion wavepacket and a photon wavepacket, the latter being subsequently transported out of the magnetosphere along geodesics determined by the photon dispersion relation in the strongly magnetised plasma. Note in the present work we include simultaneously the effects of gravity (by incorporating the curved spacetime metric of the neutron star) and strongly magnetised fields. This results in a covariant dispersion relation for photons in a magnetised plasma which have a dispersion relation~\cite{PhysRevE.64.027401,Turimov:2018ttf}
\begin{align}
    &{\cal D}(k) = g^{\mu \nu} k_\mu k_\nu \nonumber \\
    &- (\omega^2 -k_\parallel^2 ) \sum_s \frac{4 \pi q_s^2 n_s}{\gamma_s^2[\mu_s (\omega  - k_\parallel)^2 - c_s^2(\omega v_s - k_\parallel )^2]}.
\end{align}
Here, the sum $s$ is over different charge carrier species,  $\gamma_s$ is a generalised Lorentz factor, and $k_\parallel$ gives the 4-momentum projected onto the magnetic field and $v_s$ corresponds to the velocity of charge carriers and $\mu_s$ the energy per particle. The number density of each species $s$ is given by $n_s$ and the charge by $q_s$. Full definitions and detailed explanations of the various terms can be found in \cite{PhysRevE.64.027401}. We will take the non-relativistic limit, setting $v_s =0$, $\gamma =1$, $c_s=0$ and $\mu_s = m_s$. We also consider a purely electron-positron plasma so that $q_s = e$. Setting $D(k) =0$ then gives the dispersion relations for photons in a non-relativistic plasma. The equations of motion for the photon rays are then given by Hamilton's equations
\begin{equation}\label{eq:Hamilton}
\frac{d x^\mu}{d \lambda} = \frac{\partial {\cal D}}{\partial k_\mu}, \qquad  \frac{d k_\mu}{d \lambda} = - \frac{\partial {\cal D}}{\partial x^\mu}.
\end{equation}
To compute the power, we back-trace from the observer to the point of emission, following the equations of motion \eqref{eq:Hamilton}. This is analogous to the procedure used in \cite{Leroy:2019ghm,Battye:2021xvt}.

In addition, we also now include multiple axion-photon conversions arising from multiple reflections off the critical surface, as happens within ``throats" of the plasma distribution around the neutron star. These throats are partially enclosed regions near the charge separation gap off whose walls the photon can be multiply-reflected due to plasma gradients. This can enhance the power of the signal relative to not including such effects as was done in \cite{Leroy:2019ghm,Battye:2021xvt}. An extensive analysis of ray-tracing techniques which combines the physical effects considered across \cite{Battye:2021xvt} and \cite{Witte:2021arp} is currently underway and will appear in a companion paper~\cite{McDonaldInPrep}, where the full details of our scheme will be presented. This will include a systematic study of anisotropic plasmas. We do not consider the effects of so-called ``de-phasing" conjectured in  \cite{Witte:2021arp}  which awaits a more robust physical description to see if the effect persists under more mathematically rigorous formulation. We do not need to consider non-linear effects arising from very large conversion probabilities where photons may convert back into axions. This is safe for PSR J2144$-$3933 on which we performed our observations, which has sufficiently low magnetic fields that the conversion probability remains small. 

\subsection{Signal templates}

\begin{figure}
\centering
 \includegraphics[width = 0.45\textwidth]{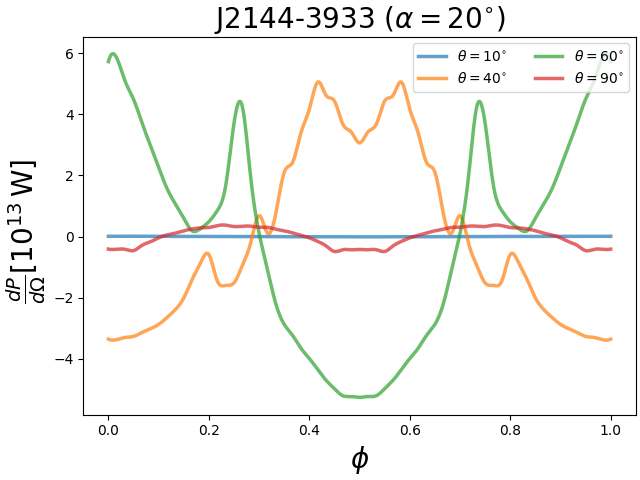}
 \caption{Time-dependence of radio templates from axion dark matter as a function of pulse phase $\phi$ for different values of $\theta$ with fixed value of $\alpha=20^{\circ}$ with $g_{a \gamma \gamma }= 10^{-10} {\rm GeV}^{-1}$.  As expected there is little time variation for $\theta=10^{\circ}$, but it can be substantial for intermediate angles. When $\theta=90^{\circ}$ the ``throats" of the GJ model never cross the line-of-sight so, although there is some emission, it is relatively weak compared to cases where they are visible to the observer.}
 \label{fig:profiles}
\end{figure}

Having outlined the basic details of our ray-tracing scheme. This can now be used to begin deriving signal templates.  In particular, these simulations allow one to model the radio signal as a function of pulsar and axion input parameters. In particular, one can compute the profile of the signal in frequency and time. The frequency dependence of the profile is determined by the mass of the axion - which sets the central frequency of the radio line. The width is set by a combination of the velocity dispersion of dark matter and by line-broadening induced by the time-dependent nature of the plasma, which modifies photon frequencies as they move through the magnetosphere. The time variation of the signal arises from the fact that the plasma surrounding the neutron star is not axisymmetric. In the present approximation we assume an electron-positron plasma which co-rotates with the star with regions of positive and negative charge separated according to the Goldreich-Julian density~\cite{goldreich1969}
\begin{align}\label{eq:GJDensity}
n_{\rm GJ} = \frac{B_0 \Omega}{2 \, e} \left( \frac{R}{r}\right)^3\Big[ &
\cos \alpha + 3 \cos \alpha \cos(2 \theta) \nonumber \\
&+ 3 \sin \alpha \cos (\phi - \Omega t)  \sin 2 \theta
\Big]\,,
\end{align}
with the plasma frequency given by $\omega_{\rm p} = \sqrt{4\pi |n_{\rm GJ}|/m_e}$.
Here, $\Omega$ is the frequency of the pulsar, $\alpha$ is the angle between the magnetic axis of the co-rotating dipole and the rotation axis of the star, $R$ is the stellar radius, and $B_0$ is the magnetic field strength on the surface at the magnetic poles. The polar coordinate $\theta$ and azimuthal angle $\phi$ are defined with respect to the rotational axis of the star. It is obvious that whenever $\alpha \neq 0$, the plasma is time-dependent with respect to a non-rotating observer. This results in time-dependent radio signals, as illustrated in Fig. \ref{fig:profiles}. For a given pulsar, the remaining input parameters to determine the axion dark matter radio signal are then the distance to the pulsar $D$, and the dark matter density $\rho_{\rm DM}$ at the position of the pulsar.

In our analysis, we will take $P$ and $B_0$ to be their quoted measured values. In principle there are some extra uncertainties which would need to be taken into account. Pulsar periods $P$ are one of the best-measured quantities in astronomy, while the magnetic field strength of the pulsar at the pole is inferred from measurements of $P$ and $\dot{P}$ the spin-down rate of the pulsar combined with model-dependent parameters including the moment of inertia of the pulsar and its radius~\cite{Belvedere2015}. This calculation is standard but assumes that the energy released by the pulsar in the form of radio emission comes from the loss of rotational energy calculated from the spin-down rate. The comparatively large values of $P\dot{P}$ observed for magnetars form the basis for their large inferred values of $B_0$, but it is also known that the magnetars are known to emit large X-ray fluxes whose luminosity cannot be explained by spin-down alone. The emission mechanism for some magnetars may be powered magnetically in contrast to the rotation powered radio pulsars \cite{Thompson_Duncan_1995, Thompson_Duncan_2001}. In particular, magnetars can have a significant toroidal component to their field whose magnitude one can constrain by imposing that the luminosity from magnetic dissipation exceeds that from the traditional spin-down. Based on observations of the X-ray flares/bursts from magnetars, the typical ratio of the toroidal to poloidal component of the field is of the order $\sim 10$ \cite{Mondal_2021, Pons_2011}. While it is possible to calculate this enhancement from running magneto-hydrodynamic (MHD) simulations, the impact on the final constraints is suppressed by $\sqrt{S} \propto B^{0.4}$ (see sec.~\ref{sec:future}) and s such, we leave a more sophisticated modelling of the magnetosphere for future work.

The distance to the pulsar, $D$ can be inferred from parallax measurements or from the dispersion of the pulse as a function of frequency, since the photons emitted in the main beam of the pulsar traverse through the galactic electron density along the line-of-sight. Given a model for the galactic electron density, one can estimate the distance to a pulsar. If we use  the pyGEDM code and the galactic coordinates as measured in the catalogue we obtain $D\approx 0.29\,{\rm kpc}$ \cite{2020ApJ...895L..49P, 2020ApJ...888..105Y, 2017ApJ...835...29Y, 2003astro.ph..1598C, 2003ApJS..146..407F, 2002astro.ph..7156C}. This illustrates that this computation relies on the underlying model for the galactic electron density distribution. A parallax measurement for this pulsar yields $D=0.165^{+0.017}_{-0.014}\,{\rm kpc}$~\cite{2009ApJ...701.1243D}. This is a more direct measurement and we will use it as our fiducial value when establishing limits, but we note that the limit is $\propto D$. Galactic dark matter profiles allow one to predict the dark matter density at the position of the pulsar, but these models become highly uncertain as one gets closer to the galactic centre, where some models predict a spike in the density, while others predict a more cored profile.

Based on these arguments, we conclude that the completely unknown quantities which parameterise the signal templates are $(\alpha, \theta)$. We, therefore, generate a simulated database\footnote{Formally we generate templates for discrete $(\alpha_i,\theta_i)$ and numerically interpolate to generate a template for a continuous range of $\alpha$ and $
\theta$. See appendix A for a description of this procedure and an illustration of its accuracy.} of periodic flux profiles as a function of $(\alpha, \theta)$. Some of these profiles are displayed in Fig.~\ref{fig:profiles} which indicates the $(\alpha,\theta)$ dependence of the time-variability of the signal. 

 In the next section, we describe how to harness the information and the larger time-variability of the signal to improve the prospects of detecting axion dark matter. 
 
\section{Searching for time-dependent signals}\label{sec:timeDomain}

In order to search for time dependent signals we will employ a matched-filter template-fitting approach similar to that used in Gravitational Wave Astronomy to detect the waveforms of the late stages of binary black hole inspirals \cite{MeyerChristensen,Allen:2004gu}. In that case, once a detection of gravitational waves was made, this allows estimates of physical parameters such as the black hole masses. Formally, if an axion were detected, one could use axion radio signals to fit model parameters of the pulsar magnetosphere. However, our ambition at this stage is much more conservative: we will use the signal-to-noise estimate from the matched filter, ${\hat q}$, defined below, to quantify the likelihood of detection. In this sense, ${\hat q}$ acts as a statistical test for whether the data is distinguishable from noise. Values of  $\hat{q}$ above a threshold then constitute a detection. Conversely, for values below this, by injecting would-be signals into the data, we obtain the expected value $q_{\rm exp}  = <\hat{q}>$ (see eq.~\eqref{eq:qexp}) from an axion signal. By comparing this to the measured value $\hat{q}$, we can exclude regions of axion parameter space.

This procedure provides a means to derive limits on $g_{a\gamma\gamma}$ as a function of $m_{\rm a}$, and importantly allows us to take into account that for a fixed value of $m_{\rm a}$ there are a wide range of templates for the expected signal due to the parameters of the particular neutron star system under consideration. These are the period of the pulsar, $P$, its surface magnetic field flux density, $B_0$, the radius of the neutron star, $R$, the angle $\alpha$ between the magnetic axis and spin axis of the star, and the angle $\theta$ between the line of sight and the spin axis. As in previous attempts to derive constraints on $g_{a\gamma\gamma}$ using neutron stars~\cite{Foster:2020pgt,Darling:2020plz,Darling:2020uyo,Battye2022,FosterSETI2022} for simplicity, as a demonstration of the filter, we do not, for instance consider uncertainties in $B_0$ or $R$ (the period, $P$ is of course measured with tremendous accuracy). We leave a computationally intensive parameter scan for future work, but this would be a straightforward extension of the existing framework.

Instead, our main focus here is on the sensitivity of the time-dependence of the signal to pulsar parameters, which is especially sensitive to the value of 
$\alpha$ and $\theta$. For each value of $m_{\rm a}$, we therefore obtain a constraint on the value of $g_{a \gamma \gamma}$ for every pair ($\alpha$,$\theta$). We can then exclude certain ranges of $\alpha$ and $\theta$ with further modelling and observations of the pulsar signal, notably the pulse width. We then use the value of $(\alpha,\theta)$ from this remaining subset which gives the most conservative constraints on $g_{a \gamma \gamma}$. 

\subsection{Derivation of matched filter and mathematical properties} \label{sec:mfmath}

Matched filters are a standard technique in signal processing and they are often used in astronomy to search for signals with a known, or parameterizable profile. This can be done in the spatial, frequency or time domains. In this section, we will derive the standard matched filter before discussing some of its properties. In section~\ref{sec:mfadv} we use it to quantify the pros and cons of time domain observations and in section~\ref{sec:mfdemo} we will show how it can be used to recover an injected signal in simulated radio data. 

There are a number of ways of formulating the matched filter. Here, we will use a discrete matched filter in which the data is represented by a vector of finite length. This can be generalised to continuous functions in which the vector inner products become convolution integrals over functions (see \cite{Allen:2004gu}). The discrete formulation has the advantage of simplifying notation. 

Our starting point is the so-called  ``data vector'', ${\bf d}$.
We will assume the data is the sum of a signal $\textbf{S} = S_0 \textbf{F}(\textbf{p})$ and some additive noise $\hat{\textbf{n}}$
\begin{equation}
    \textbf{d} = S_0 \textbf{F}(\textbf{p}) + \hat{\textbf{n}} \, .
\label{SNR}
\end{equation}
Here, we have decomposed the signal according to 
\begin{equation}\label{eq:Fnorm}
|{\bf F}| \equiv  \sqrt{\textbf{F}\cdot\textbf{F}} = 1
\end{equation}
 so that $S_0$, gives the root mean squared flux density of the signal 
 \begin{equation}\label{eq:S0Definition}
     S_0 = \sqrt{\textbf{S}\cdot\textbf{S}}\,. 
 \end{equation}
 The signal is further characterised by a ``parameter vector" ${\bf p}$.  In section~\ref{sec:SignalProperties} we will compute a number of templates ${\bf F}({\bf p})$ for the signal where ${\bf p}=(m_{\rm a},\alpha,\theta)$. 
The noise vector $\hat{\textbf{n}}$ is assumed to be Gaussian with $\langle{\bf n}\rangle=0$ and $\langle {\bf n}{\bf n}^{T}\rangle=C$ where $\langle..\rangle$ denotes an ensemble average of noise realizations and $C_{ij}$ is the covariance matrix with $i,j=1,..,n_{\rm d}$ where $n_d$ is the total number of data points. 

Assuming a Gaussian likelihood, $-2\log{\cal L}=\chi^2$, one can calculate the maximum likelihood estimate ${\hat S}_0$ by minimizing 
\begin{equation}
\chi^2=({\bf d}-S_0{\bf F})^TC^{-1}({\bf d}-S_0{\bf F})\,.
\end{equation}
In the above equation, we assume the data vector $\textbf{d}$ contains the true signal $\textbf{S} = \textbf{S}_{\rm true}$ so that minimising $\chi^2$ above can be thought of as minimising a generalised least-squares difference (relative to the noise in each channel - hence the factor $C^{-1}$) between the true signal and possible templates $S_0 \textbf{F}$.  Viewing $\chi^2$ as an unknown function of $S_0$, we can find the minimising value, $\hat{S}_0$ given by
\begin{equation}
    {\hat S}_0 = \frac{\textbf{F}^{T} C^{-1} \textbf{d}}{\textbf{F}^{T} C^{-1} \textbf{F}}\,.
\end{equation} 
 One can also deduce the ``matched filter noise"
\begin{equation}
    \sigma_{\rm MF}= \left(\textbf{F}^{T} C^{-1} \textbf{F}\right)^{-1/2} 
\end{equation}
and the  signal-to-noise estimate ${\hat q}$ may then be written as 
\begin{equation}
    {\hat q} = \frac{{\hat S}_0}{\sigma_{\rm MF}} = \frac{\textbf{F}^{T} C^{-1} \textbf{d}}{\left(\textbf{F}^{T} C^{-1} \textbf{F}\right)^{1/2}}\, .
\end{equation}
It is important to understand the difference between the noise in the data, characterised by $C$,  and the ``matched filter noise", $\sigma_{\rm MF}$. They are related, but $\sigma_{\rm MF}$ also depends on the filter. We return to this issue in the next section.

In order to understand properties of the matched filter we will assume diagonal covariance matrix 
\begin{equation}\label{eq:covariance}
C_{ij}=\sigma_{N}^2\delta_{ij}
\end{equation}
where $\delta_{ij}$ is the Kronecker delta and $\sigma_{N}$ is the noise in each channel - all of what said here can be adapted to the case of a general covariance matrix, but it is less simple to see. 

In general, the data has a number of dimensions, for example, space, time and/or frequency, then $\textbf{d}$ has $n_{\rm d} = n_1\times ... \times n_k$ entries where $n_i$ for $i=1,...,k$ are the number of points in each of the dimensions. In our case we will search in the frequency and time directions so the number of entries in the data vector will be $n_{\rm d} = n_{\rm f}\times n_{\rm t}$ where $n_{\rm f}$ is the number of frequency channels and $n_{\rm t}$ is the number of time samples. In that case, the data vector could be written as
\begin{equation}\label{eq:datvec}
        \textbf{d} = \Big( d^{\omega_1}_{t_1},\dots, d^{\omega_1}_{t_{n_t}}, \dots ,d^{\omega_{n_{\rm f}}}_{t_1},\dots, d^{\omega_{n_{\rm f}}}_{t_{n_t}}\Big),
\end{equation}
where $d^{\omega_i}_{t_j}$ labels the data vector in the $i{\rm th}$ frequency bin and $j{\rm th}$ time-channel. The covariance matrix of the form \eqref{eq:covariance} is then nothing more than the statement that the noise between all possible pairs of time and frequency bins is totally uncorrelated. 

Returning to our main discussion, it follows that for a covariance matrix of the form \eqref{eq:covariance}, we have
\begin{equation}\label{eq:FilterTest}
    {\hat q}= \frac{{\bf F}\cdot{\bf d} }{\sigma_{N}}\,,
\end{equation}
that is, the dot product of the filter $\textbf{F}$ with the data vector. When $\textbf{F}$ matches that in the true signal, we have $\langle{\hat q}\rangle=S_0/\sigma_{N}$ (from eq. \eqref{SNR}).

Now assume that ${\bf p}_{\rm true}$ has entries which are the true parameters. The ensemble average of ${\hat q}$ for a filter with arbitrary parameter, ${\bf p}$ is
\begin{equation}\label{eq:qexp}
    \langle {\hat q}\rangle={\bf F}({\bf p})\cdot {\bf F}({\bf p}_{\rm true}){S_0\over\sigma_{N}}\,. 
\end{equation}
Next, we note that as a trivial consequence of the Cauchy-Schwarz inequality, we have ${\bf F}({\bf p} )\cdot {\bf F}({\bf p}_{\rm true}) \leq \left| {\bf F}({\bf p} ) \right| \left| {\bf F}({\bf p}_{\rm true})  \right|$ with equality if and only if ${\bf F}({\bf p} ) = {\bf F}({\bf p}_{\rm true})$. Assuming that the template is non-degenerate with respect to the values of $\textbf{p}$,  this occurs only for  ${\bf p}={\bf p}_{\rm true}$ for which ${\hat q}$ is then maximal. Thus $\hat{q}$ acts as a likelihood test for the values of $\textbf{p}$.

In what follows, since the line width of the axion is typically less than the width of our frequency channels, the vector \eqref{eq:datvec} is sparse for a given value of $m_a$, with non-vanishing entries in only one frequency channel where $\omega = m_a$. This means filters with different values of $m_{\rm a}$ are orthogonal. More generally, with higher frequency resolution, we would expect to be able to probe both the time and frequency structure of the signal.  By contrast, filters with different values of $\alpha$ and $\theta$ are not orthogonal and will in general have overlap such that $\textbf{d}(\theta, \alpha , m_a)\cdot \textbf{d}(\theta', \alpha', m_a) \neq 0$.  

In principle, this means  an axion detection would allow us to determine likely values of the pulsar parameters in analogy to the way in which observable gravitational wave signals allow inference of the mass and spin of their associated black holes. However, our present approach will be to minimise the expected signal over a conservative subset of values of $(\alpha,\theta)$, thereby obtaining the most conservative constraints on $g_{a \gamma \gamma}$ for allowed values of angles.

In order to turn our continuous axion signals into discrete vectors we must perform some kind of coarse-graining. We therefore define a binning scheme for $N$ time-channels centered on the points $t_i = (i-1/2)\Delta t $ where $i = 1,..,N$ and $\Delta t = 1/N$ so that the discretized signal is given by
\begin{equation}
    S^{\omega}_{t_i} = \frac{1}{\Delta t} \int^{t_i+ \Delta t/2}_{t_i - \Delta t/2} dt S(t,\omega),
\end{equation}
where $S(t,\omega )$ is the flux density of the axion signal as a function of time (and frequency) that we derive using our radio signal models.  

\subsection{Why do a time domain analysis?}
\label{sec:mfadv}
In this section we will discuss the advantages of doing a time domain analysis in terms of increasing the chances of detecting axions. We will also discuss the issue of whether subtraction of the pulse-average from the time-domain data (as is often the case in pulsar observations and as we have done in our data) will have significant impact. 

In order to do this we will investigate some properties of the matched filter. Consider applying the matched filter to a given frequency channel whose signal consists of $N$ time-channels is 
\begin{equation}
    \textbf{S} = (S_1 , \dots,  S_N).
\end{equation}
Then according to Eqs \eqref{SNR}-\eqref{eq:S0Definition} and \eqref{eq:FilterTest}, the matched filter will return
\begin{equation}\label{eq:qtime}
    \braket{\hat{q}} = \frac{\sqrt{\textbf{S}\cdot\textbf{S}}}{\sigma_{N}} ,
\end{equation}
where $\sigma_{N}$ is the noise on each of the $n_t = N$ time-channels. The noise amplitude $\bar{\sigma_{N}}$ averaged across all times is then given by $\bar{\sigma} =\sigma_{N}/\sqrt{N}$. We can therefore re-write \eqref{eq:qtime} as
\begin{equation}\label{eq:qtdep}
    \braket{\hat{q}}_{\rm time} = \frac{\left(\sigma_S^2 + \mu_S^2 \right)^{1/2}}{\bar{\sigma}}
\end{equation}
where $\mu_S = \Sigma_i S_i/N$  and $\sigma_S = \sqrt{\Sigma_i S_i^2/N - \mu^2_S}$ are the average and standard deviation of $\textbf{S}$, respectively. 

Now let us consider carrying out a measurement with no time resolution. This is the case when one simply uses the telescope to make a total flux measurement over a long observing time. In this case the noise is again $\bar{\sigma}$ given by averaging the noise over all integration time, the signal $\sqrt{\textbf{S}\cdot \textbf{S}}$ is simply given by the mean $\mu_S$ so that
\begin{equation}\label{eq:qtIndep}
    \braket{\hat{q}}_{\rm flux-avg.} = \frac{\mu_S}{\bar{\sigma}}.
\end{equation}
This can be thought of as a trivial matched filter with a single time channel, in which any fine-grained time information has been lost. This is in effect how all previous single pulsar observations for axion dark matter have been carried out~\cite{Darling:2020plz,Darling:2020uyo,Battye2022}.

By comparing the cases of a time-domain analysis \eqref{eq:qtdep} with a total-flux measurement \eqref{eq:qtIndep} it becomes immediately apparent that since $\sigma^2_S \geq 0$, the time-domain analyses will always equal or outperform the time-averaged measurement. Thus time-domain information increases the potential to detect axions. In particular, when the relative time-variation is large ($\sigma_S/\mu_S \gg 1$), the time-domain search provides a gain in sensitivity by increasing the signal to noise by a factor $\simeq \sqrt{\sigma_S/\mu_S}$, i.e. the square-root of the relative variance. This is a simple consequence of the fact that $\hat{q}$ is proportional to the root-square of the signal, and so it implicitly encodes information about its variability.  The same observation was also made in \cite{Leroy:2019ghm} but the matched filter allows this to be justified from first-principles. 

Although this is in general not necessary, the observations used later in the paper have had the average of the signal subtracted~\footnote{Observations of pulsars are often taken in this way since they are probing the time-variable signal from the rotation of the neutron star. Ultimately, there is nothing to prevent the use of the averaged signal as well, but it can require extra work to calibrate it. Here, we see that if the expected time variation is significant, removing the average does not significantly affect sensitivity to axions.}. Thus while we retain time-domain information, the baseline $\mu_S$ of the signal is essentially re-normalised to zero. In this case, the time-domain analysis gives a baseline subtracted (BS) value
\begin{equation}\label{eq:BaselineSub_SNR}
     \braket{\hat{q}}_{\rm time-BS} = \frac{\sigma_S }{\bar{\sigma}} .
\end{equation}
Clearly, when there is only a small time variation, ($\mu_S \gg \sigma_S$), subtracting the baseline leads to a lower value of $\braket{\hat{q}}$, relative to retaining it as in eq. \eqref{eq:qtdep}. This is for the simple reason that if the average signal is large, one loses a lot of signal power by removing the average. Conversely, in the regime where there is large time variation, $\mu_S \ll \sigma_S$ the time-domain analysis \textit{with} baseline subtraction performs almost as well as eq.~\eqref{eq:qtdep}. This is again the statement that high peaks above the noise in the time-domain still allow for good discrimination from noise. 

\subsection{Demonstration on simulated data}
\label{sec:mfdemo}
In the previous subsections we have derived the main properties of the matched filter estimate of the signal-to-noise ratio. This is not particularly new to those with a background in astronomy, but may not be familiar to a general reader. The matched filter is the optimal filter for a Gaussian likelihood and would be the clearest way to identify a signal in the data with ${\hat q}$ being a proxy for the signal to noise of detection.

In this subsection, to aid understanding, we will demonstrate the performance of the matched filter in a toy example by injecting a signal for a neutron star with the same physical characteristics as PSR J2144$-$3933 into simulated data with similar noise to the observations that we have in hand, described in sections~\ref{sec:res_tech} and \ref{sec:res_noise}. 

\begin{figure}
 \includegraphics[width=0.45\textwidth]{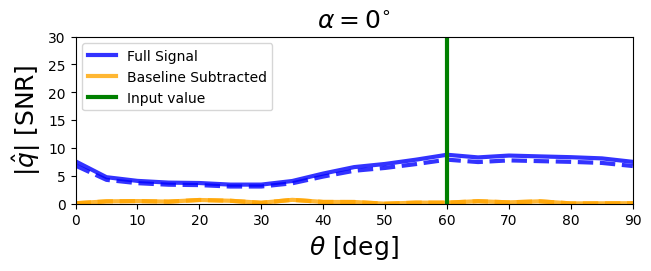}
 \includegraphics[width=0.45\textwidth]{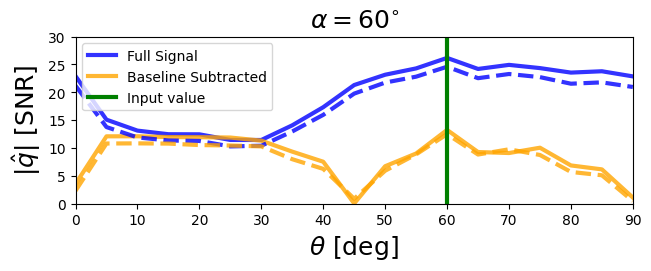}
    \caption{An illustration of using the matched filter SNR to search for the signal. $\hat{q}$ is shown as function of observing angle $\theta$ returned by the matched filter \eqref{eq:FilterTest} with simulated Gaussian noise similar to that expected for the observations of PSR J2144$-$3933 discussed in this paper. We display the SNR for two scenarios for the input signal, one with $(\alpha,\theta)=(0^{\circ},60^{\circ})$ (case A, top panel) and another with $(\alpha,\theta)=(40^{\circ},60^{\circ})$ (case B, bottom panel). In both cases we choose a noise amplitude of $\sim 2.5\,{\rm mJy}$ consistent with that observed in our data for $m_{\rm a} = 4.2\,\mu{\rm eV}$. We use an axion mass of $3.7\,\mu{\rm eV}$ and $g_{\rm a\gamma\gamma} = 10^{-10}\,{\rm GeV^{-1}}$. As an additional sanity check, we repeat the analysis by inserting the signal into the real pulse-subtracted data (see sec.~\ref{sec:Observations} for details on the data), represented by the dotted lines. The fact these lines appear to agree indicates that our noise model is representative of the situation in the real data.}
    \label{fig:MF_demo_sim}
\end{figure}

We will inject signals with a mass of $m_{\rm a}=4.2\,\mu{\rm eV}$ which corresponds to an observing frequency $f_{\rm obs} \simeq 1.0{\rm GHz}$ and $g_{a\gamma\gamma}= 10^{-10}{\rm GeV}^{-1}$. We will consider two different choices for the angles $\alpha$ and $\theta$. Case A with $(\alpha,\theta)=(0^{\circ},60^{\circ})$ is close to an aligned rotator and hence we would expect no time variation, whereas case B with $(\alpha,\theta)=(40^{\circ},60^{\circ})$ has a very strong time variation.
We will consider two observations: one in which the pulse-averaged signal power is retained, and another where it is removed which is more common in pulsar observations as we have explained earlier. We have already discussed the pros and cons of the two approaches in section~\ref{sec:mfadv} and this is just an illustration of the specific point. The full results of these test cases are shown in Fig.~\ref{fig:MF_demo_sim}. 

For a nearly aligned rotator ($\alpha = 0^\circ$), the pulsar is axisymmetric about its rotation axis, and the signal has no time-dependence. Therefore, in this case, if one searches for the signal with the pulse-average removed, there is by definition no signal present in the effective data vector, leading to a non-detection. The filter is essentially scanning a particular noise realisation with zero signal. In the bottom panel, the input signal contains significant time-dependence (roughly an order of magnitude). Therefore, the input signal is detected with a SNR of the same order of magnitude as in the total power case, in accordance with the discussion comparing \eqref{eq:qtdep} and \eqref{eq:qtIndep}. In all the cases except the baseline-subtracted $\alpha=0$ case, the filter successfully returns the maximal SNR for the input value of $\theta$. 

\section{Observations of PSR J2144-3933 with MeerKAT}\label{sec:Observations}
\label{sec:res_tech}

In order to test this idea we selected PSR J2144$-$3933. We did this by considering the list of observed neutron stars~\cite{Manchester:2004bp}\footnote{\url{https://www.atnf.csiro.au/research/pulsar/psrcat/}} which provides estimates for $B_0$, $P$ and the pulsar distance $D$. In order to make an estimate of the strength of the signal expected for a particular pulsar we use an analytic formula based on the radial trajectories approach~\cite{hook2018}. Although this assumption has been shown to be not sufficiently correct to provide accurate predictions~\cite{Battye:2021xvt,Witte:2021arp} it is likely that it gives a reasonable figure of merit since it should have the correct scaling with the important parameters. We will discuss the issue of what is the optimal target in more detail again in section~\ref{sec:future} in light of what we have learnt. Specifically, we have used the figure of merit 
\begin{equation}\label{eq:FOM}
\hbox{FOM}=\rho_{\rm DM}{B_0^{2/3}P^{7/3}\over D^2}\,,
\end{equation}
where $\rho_{\rm DM}$ is the density of dark matter expected in the vicinity of the pulsar, to create a ranked list of pulsars. This formula can be derived from results presented in \cite{hook2018,Battye_2020}

We presume that all the dark matter in the Galactic halo is in the form of axions, the standard assumption when obtaining limits, and  extrapolate the local density of $\rho_{\rm DM}\approx 0.45\,{\rm GeV}\,{\rm cm}^{-3}$ using an NFW profile for dark matter in the galaxy. Except in the very centre, near the location of the Galactic Centre Magnetar (GCM) PSR J1745$-$2900\footnote{Most attempts to constrain axions using neutron stars have used the GCM as their target, attracted by the large magnetic field, and indeed we re-visit using it for this type of analysis in section~\ref{sec:future}. Depending on the parameters of the NFW profile we used it varied from around 10th in our list to 1st. A clear reason to not use it is that there is an additional uncertainty created by the lack of knowledge of the dark matter density in the centre of galaxy.}, this is likely to give a reasonable estimate of the trade-off between $\rho_{\rm DM}$ and $D$ in the FOM.

The PSR J2144$-$3933 which has $B_0\approx 2.1\times 10^{12}\,{\rm G}$ estimated from the $P$ and ${\dot P}$ based on  electromagnetic spin down, $P=8.51\,{\rm sec}$ and $D=0.12\,{\rm kpc}$ came third on the list \footnote{The objects PSR J0736$-$6394 and PSR J1856$-$3754 came first and second on the list. The first is Rotating RAdio Transient (RRAT) which is not monitored routinely in the radio waveband, while the second has only been detected in the X-ray waveband. Both these objects are  unsuitable for our study here.} and seems an ideal object. This object has a long period, and hence a strong axion signal, but is otherwise unremarkable. The fact that it is very nearby is also a significant advantage since it means that we can be more sure about the local value of $\rho_{\rm a}$ used in our predictions. 

Within the GJ model there is a maximum axion mass~\cite{Battye2022} given by 
\begin{eqnarray}
    m_{\rm a}^{\rm max}\approx 85\,\mu{\rm eV}\left({B_0\over 10^{14}{\rm G}}\right)^{1/2}\left({P\over 1\,{\rm sec}}\right)^{-1/2}\cr\times \left(1+{1\over 3}\cos\alpha\right)^{1/2}\,,
\end{eqnarray}
which is $\approx 4.7\,\mu{\rm eV}$ corresponding to $f_{\rm obs}\approx 1.15\,{\rm GHz}$ for this object. It is not possible to use any observations above this frequency in obtaining a limit using the GJ model  predictions, but we do use the data in our search for generalised periodic signals in section~\ref{sec:periodic}.

The specific observation of PSR J2144$-$3933 used in this work was taken at 2020-07-13 02:20:47 as part of the MeerTime Large Survey Programme on MeerKAT. The observation was recorded as part of the Thousand Pulsar Array~\cite{Johnston2020} census observations and hence used the `full' MeerKAT array, specifically in this instance 58 of the antennas were used to form a single tied array beam pointed at the pulsar.
For long-period pulsars the Thousand Pulsar Array census aims to record $\gtrsim 512$ pulses from each pulsar, and hence the total observing duration was $4416\,\mathrm{s}$, much longer than typical Thousand Pulsar Array observations.
The data produced by the MeerKAT beamformer are processed in real time by the PTUSE instrument~\cite{Bailes2020}, folding the data with the known period of the pulsar.
Post processing, including initial automated cleaning of radio frequency interference and flux calibration is carried out on the Swinburne OzStar supercomputer using the MeerPipe pipeline developed by MeerTime.
The calibration and cleaning procedure used for the Thousand Pulsar Array data is described in \cite{Posselt}. The output data have 1024 rotational phase bins and 928 frequency channels, each of width $0.8359375\,$MHz (total bandwidth $775.75\,$MHz), and centred at $1283.58203125\,$MHz.

\subsection{Modelling the noise}
\label{sec:res_noise}

\begin{figure}
    \centering
    \includegraphics[scale=0.5]{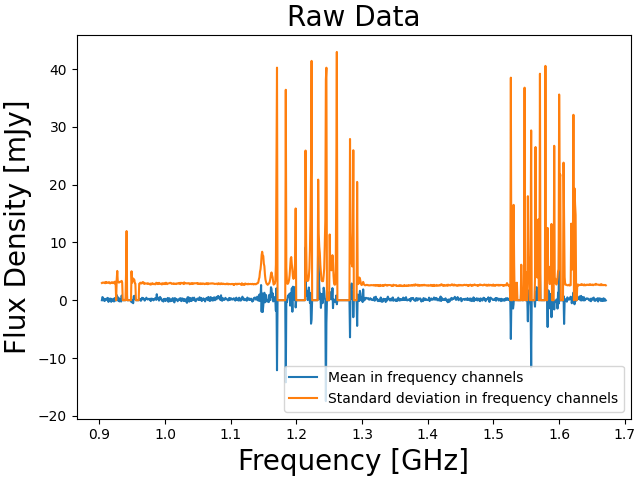}
    \medskip\medskip
    \includegraphics[scale=0.5]{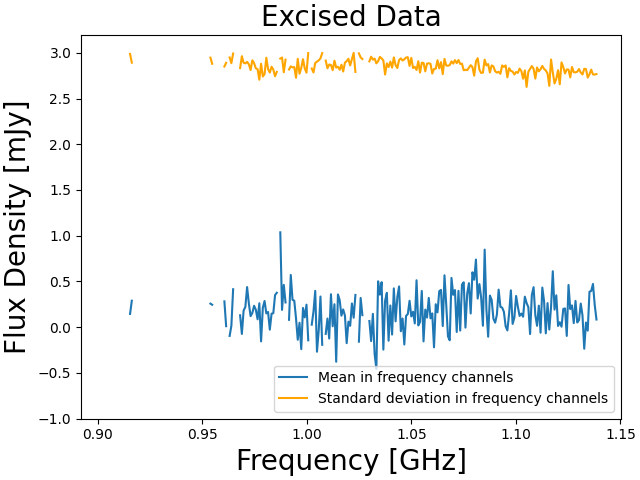}
    \caption{Measurements of the mean, $\mu_S$ and $\sigma_S$ standard deviations of the observed off-pulse flux density of PSR J2144$-$3933 used in this paper. The average is over pulsar phase for a fixed frequency. In the top panel, we show the $\sigma_S$ and $\mu_S$ for the full data-set. In the bottom panel, we show the equivalent after employing a cut of $\sigma_N=3.0\,{\rm mJy}$ but with a different scale. As demonstrated in the figure this cut allows us to excise the frequency channels that are dominated by RFI contamination, with the remaining channels being compatible with $\mu\approx 0$ and $\sigma_N\approx 2.8\,{\rm mJy}$. 
    }
    \label{fig:data_frequency_channels}
\end{figure}

\begin{figure}
    \includegraphics[scale=0.5]{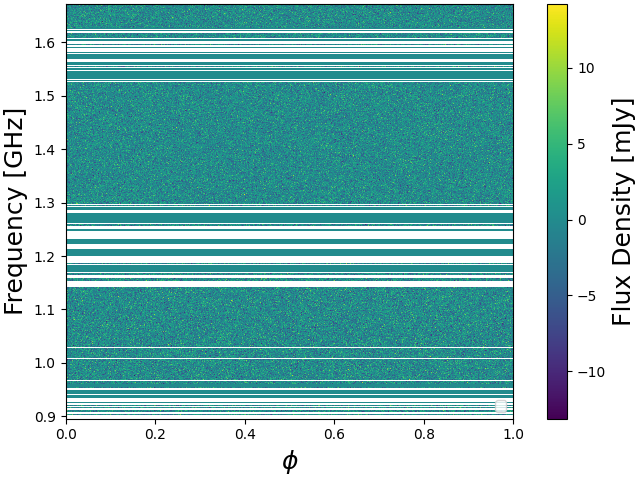}
    \medskip\medskip
    \includegraphics[scale=0.5]
    {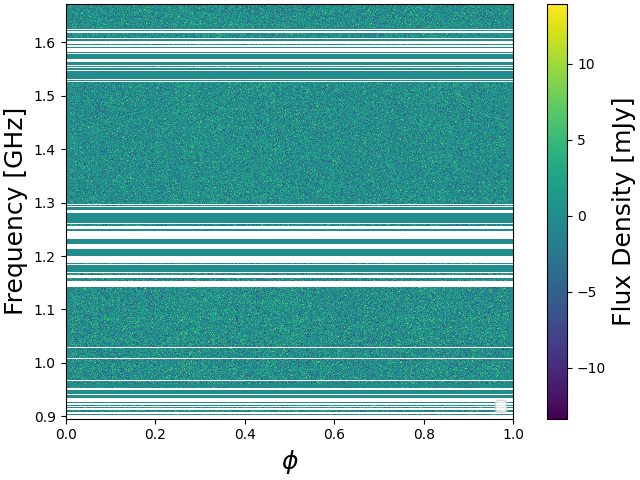}
    \caption{Dynamical spectra (i.e. the data as function of frequency and pulsar phase) with the RFI dominated channels excised and also the pulse (around $\phi=0$ and $\phi=1$) removed. The top panel is our noise model described in the text, while the bottom panel is the actual data. 
    }
    \label{fig:cut_data}
\end{figure}

The observed data has already been processed to remove the effects of Radio Frequency Interference (RFI). This is sufficient to locate the main peak of the pulsar pulse, which is typically much stronger than the axion signal. The first thing we do is remove the pulsar main beam signal from the time-domain, so that the remaining data is in the off-phase of the pulsar. We do this by excising 20 time channels from our data. In the top panel of Fig.~\ref{fig:data_frequency_channels}, for the remaining data we present the average over the pulsar phase for, $\mu_S$, and the standard deviation, $\sigma_S$ of the data as a function of the frequency. It is clear that, despite this procedure, there remains some low amplitude RFI in certain frequency channels and it is clear that the frequency channels affected by this must be discarded for the purposes of locating axion signals. This RFI is typically due to mobile phone, $f\sim 0.95\,{\rm GHz}$, and  Global Navigation Satellite System (GNSS), $f\sim 1.2\,{\rm GHz}$ and $f\sim 1.6\,{\rm GHz}$\footnote{We note that Karoo site where the MeerKAT telescope is situated, is one of the cleanest radio observation sites in the world and still  there is low-level RFI in these bands which would likely be extremely difficult to remove, meaning that there are some ranges of axion mass that will remain unattainable to this technique, and indeed any other searching for astrophysical axion signals.}.

This residual RFI can be removed by excising any data with $\sigma_N>3\,{\rm mJy}$  as seen Fig.~\ref{fig:data_frequency_channels} with the excised data presented using a narrower flux scale. This data appears to be relatively clean and free from obvious terrestrial RFI since it removes the regions known to be affected by known irreducible interference. Once this is done it seems reasonable to try to model the noise in the data to be an uncorrelated Gaussian random process with zero mean within each channel and standard deviation $\sigma_N(f)$ which is given by that measured in a given channel. This is a small variation - by allowing the standard deviation to vary with frequency - to the approach we have used in section~\ref{sec:mfmath}. It is unlikely that the assumption of exact Gaussianity and zero-correlations is entirely perfect, and indeed in the subsequent sections we find some evidence to suggest that there may be weak correlations in the noise as a function of the pulsar phase $\phi$. Nonetheless, we will argue that this just leads to conservative constraints and therefore, we will proceed to use this model. We note that if we are able to accurately model the correlations in the data, this could be handled by the match filter approach.

In what follows we will use this noise model to obtain limits on axion signal and, therefore, we should examine to what extent our data resembles Gaussian noise for the full dynamical spectrum, which is the term used to talk about the data as a function of frequency, $f$, and pulsar phase, $\phi$. As a self-consistency test of this noise model, we compare the real data set with that randomly generated from this distribution and this is presented in Fig.~\ref{fig:cut_data}. Visually, the two datasets appear to be very similar and on that basis we conclude that the models are compatible with each other.

\subsection{Constraining the magnetic orientation $\alpha$ and observing angle $\theta$}
\label{sec:magorientation}

We have already pointed out that the amplitude of the time dependence of the signal depends on the values of $\alpha$ and $\theta$ - this is also an issue when using the time averaged signal (see \cite{Battye2022}, for example). In particular, we have seen that there can be very little time variation when these angles are small. Therefore, we will need some further information on the pulsar geometry to enforce a constraint on $g_{a\gamma\gamma}$. 

Obtaining precise values for the pulsar geometry relies in general on strong assumptions on the observed properties of the neutron star's radio pulse (e.g. \cite{jkk+22}).
Fortunately, from the point of view of the present discussion, we  only need to rule out small angles, and when one observes a narrow pulse profile - which is the case here  - it is unlikely that the magnetic and observation axes are aligned with the spin axis. In what follows we will describe a simple model for the pulsar beam geometry with very conservative assumptions and use it in the case of PSR J2144$-$3933 to argue that one can ignore the region of parameter space around $\theta\approx \alpha\approx 0$.

Let $W$ be the pulse width corresponding to a fully illuminated circular radiation beam with half-opening angle $\rho$. These parameters can be related to the parameters in our misaligned rotator model for the neutron star $(\alpha,\theta)$ using \cite{LorimerKramerHandbook, Gil1984}
\begin{align}
\label{EqPulsarBeamRho}\cos\rho    &= \cos\alpha\cos\theta+\sin\alpha\sin\theta\cos(W/2)\, .
\end{align}
The width of the profile for PSR J2144$-$3933 is measured at 10\% of the amplitude is $(2.1\pm0.2)^\circ$~\cite{pkj+21}. It is possible that the profile is asymmetric and hence assuming that the full open field line region is active is not necessarily true. This means that the $W$ in (\ref{EqPulsarBeamRho}) should be interpreted as the pulse width that would be observed if the full beam is active. In rare cases, the middle of the open field line region is centred in between one of the profile peaks, and part of the otherwise maybe double profile is missing. Based on these two caveats, we conservatively take $W$ to be in the range $1.5^\circ$--$5.0^\circ$ for this object. 

\begin{figure}
    \centering
    \includegraphics[width = 0.45\textwidth]{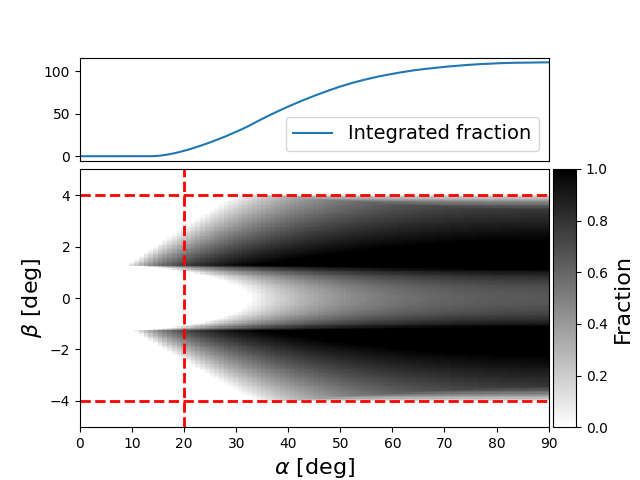}
    \caption{The constraint on the $\alpha-\beta$ plane based on the geometry of the beam where $\beta=\theta - \alpha$. As explained in the text, we define the likelihood to be the percentage  of possible values of $W$ for which solutions exist within the range of $h_{\rm em}$ we have allowed. Improbable  solutions with $\left|\beta/\rho\right| > 0.95$ are rejected. The top panel shows the vertical integration of the bottom panel and it demonstrates that solutions with $\alpha < 20^{\circ}$ are very unlikely because  they would require fine tuning of $\beta$. Therefore, we rule out $\alpha<20^{\circ}$ and $|\beta|>4^{\circ}$ as statistically unlikely, indicated by the red-dashed lines, when we calculate our limit on $g_{\rm a \gamma\gamma}$. }  
    \label{fig:theta_alpha_constraint}
\end{figure}

We now turn to the estimation of $\rho$, whose uncertainty mainly stems from a lack of knowledge of the height $h_\mathrm{em}$ at which the emission occurs. The beam is bounded by the tangents to the last open magnetic field lines. Assuming that the field is dipolar, one finds that (e.g. \cite{Rookyard2015})
\begin{align}
\rho &\approx  \frac{3}{2}\sqrt{\frac{h_\mathrm{em}}{R_{\rm c}}} = \sqrt{\frac{9\pi h_\mathrm{em}}{2cP}}.
\end{align}
where we have used the small angle approximation for $\rho$. In the second expression we have replaced the light cylinder radius $R_c$ with the pulse period $P=2\pi R_{\rm c}/c$.  Estimation of $h_{\rm em}$ is complicated by the fact that the beam is not necessarily filled, but across the pulsar population $h_{\rm em}$ at 1.4~GHz has been constrained to be in the range of $200$--$400\,{\rm km}$ irrespective of pulse period~\cite{Johnston2019}.  We take a more conservative range of $100\leq h_{\rm em} \leq 1000\,{\rm km}$, so as to ensure that we are not strongly wedded to the modelling assumptions in the pulse-beam simulations.

Note that, in principle, it is possible to compute the emission height $h_{\rm em}$ from estimates of the swing of the polarisation angle (PA) (see for example \cite{Weltevrede2008}) where theoretical PA profile is derived assuming the rotating vector model \cite{Radhakrishnan1969}. However, the detailed applicability of the rotating vector model is a debated topic in the literature and not all pulsars exhibit the characteristic banana shape that the model predicts (see for example \cite{yuen_melrose_2014}). In addition, it may not always be possible to measure the polarisation properties of candidate pulsars, particularly magnetars that do not emit at radio frequencies. We remark that while measuring $\alpha$ can significantly impact the sensitivity of our method, it remains to be seen whether this method can be considered an guaranteed way of breaking the $\alpha-\theta$ degeneracy. 

Since parameters for a given pulsar are uncertain, values of $\alpha$ all the way down to zero are allowed by (\ref{EqPulsarBeamRho}), for which there would be no time-variation. We, therefore, appeal to further arguments which allow us to place a lower bound on $\alpha$ by excluding implausible  geometries.

A problem with very small $\alpha$ geometries is that in order to explain the very narrow $W$ the line-of-sight needs to graze the very outer part of the beam such that most of the beam is invisible to us. This requires fine tuning which is not only unlikely~\cite{dyks2015}, but is also contrived for two reasons. First of all, a small change in emission height as expected for different observing frequencies~\cite{cor78} would lead to a drastic change in the observed pulse width, which is not observed~\cite{pkj+21}. Secondly, a narrow pulse not only requires a grazing line of sight, but also a circular beam with a hard edge. In reality the pulsar beam  does not have a hard edge, and hence the observed pulse shape from a grazing line of sight will be dominated by the intrinsic smoothness of the beam which will be much wider than predicted from the circular model. 

To quantify why small $\alpha$ geometries are unlikely, we construct an effective probability distribution for $\alpha$ and $\beta$ which essentially measures the number of beam realisations associated to each pair $(\alpha,\beta)$ assuming a uniform distribution for $W$. This is shown in Fig.~\ref{fig:theta_alpha_constraint}. We constructed this distribution according to the following algorithm (i) uniformly sample W between $1.5^\circ$ and $5.0^\circ$ in 100 steps. (ii) For each given W, scan over a discrete grid of $(\alpha ,\beta)$ values between $0$ and $\pi/2$. (iii) For each point $(\alpha ,\beta)$ calculate $\rho(W,\alpha,\beta)$ with Eq.~\eqref{EqPulsarBeamRho}. For each $(\alpha,\beta)$  (for the specific $W$ under consideration) if the resulting $\rho$ satisfies  $100\leq h_{\rm em} \leq 1000\,{\rm km}$ and  $|\beta/\rho| < 0.95$ record a value of 1. Otherwise assign it $0$. Note the second constraint is designed to exclude a line-of-sight with an impact parameter $\beta\simeq\rho$, which is both unlikely and implausible for the reasons above. At the end of this process, for each $W$, one has an $\alpha$-$\beta$ grid with entries that are 1 (implying an acceptable beam geometry exists) or 0. Since $W$ is sampled with 100 steps, there are 100 grids. (iv) Fig.~\ref{fig:theta_alpha_constraint} then shows the sum of these grids (appropriately normalised).

The main features that stand out are that $|\beta|$ needs to be small in order for the line of sight to intersect the beam and small $\alpha$ solutions are excluded.
No solutions exist for $\alpha\lesssim 10^\circ$. Furthermore, one can see in the top panel that solutions $\alpha\lesssim20^\circ$  are unlikely geometrical solutions, which is because they require a fine tuned (large) $\beta$. It should be stressed that $\alpha\lesssim 20^\circ$ geometries are not just unlikely, we have also argued them to be contrived\footnote{A less conservative limit on the acceptable solutions such that $|\beta/\rho| < 0.90$ would lead to $\alpha\gtrsim25^\circ$, demonstrating that the conclusions are relatively insensitive to the precise limits chosen.}. In what follows we will impose $\alpha > 20^{\circ}$ and $|\beta|<4^{\circ}$. We would expect to be able to apply similar arguments to a large fraction of pulsars that we might want to use to constrain $g_{a\gamma\gamma}$. 

\subsection{Constraints on $g_{a\gamma\gamma}$ from PSR J2144$-$3933}
 
We have shown in section \ref{sec:timeDomain} how one can compute the signal-to-noise (SNR) parameter $q$ as a function of the input parameters $(m_{\rm a}, \theta, \alpha)$ using the matched filter. In order to now derive constraints, we have carried out a parametric search for all possible profiles in our interpolated library (see Fig.~\ref{fig:profiles}) using the pulsar data presented in Fig.~\ref{fig:cut_data}. This procedure then gives us a distribution of SNR values $q_{\rm meas.}$ associated to each profile. We could repeat this process, but this time using our noise model which by definition has no signal present. Remember that it assumes uncorrelated Gaussian noise which simplifies the matched filter. This means that the values of $q_{\rm meas.}$ should be Gaussian distributed with zero mean and unit variance.

We do find that they are compatible with a Gaussian distribution that has zero mean. However, we find that the standard deviation is $\approx 1.45$ somewhat higher than the expected value which points to the fact that the noise model we are using is not optimal~\footnote{We have assumed that he noise is uncorrelated in phase which is is unlikely to be precisely true and this could easily lead to more structure than would be expected for a totally random realization of the noise. It is clear that one could achieve slightly tighter constraints by improvement of the noise model.}.  We find that, out of the $\approx 3\times 10^{4}$ templates, there are three that have $q>5$ which, if the noise model were perfect would suggest candidate detections, but in order to assess their statistical significance they should probably be scaled down by $1/1.45$ reducing them to $\approx 3.5$. This suggests that they are chance alignments with the templates; a conclusion that is further strengthened by the observation that they, and indeed the other higher values of $q_{\rm meas.}$,  appear to be randomly distributed with $m_{\rm a}$. We are satisfied that our data are compatible with a null detection.

In the case where the baseline has been subtracted from the data, which is the case for the data being considered here, the constraining power of the matched filter is determined by \eqref{eq:BaselineSub_SNR}. For the pulsar we have chosen, Fig.~\ref{fig:profiles} shows the time-dependence of the profiles. Note the relative variance vanishes at $\alpha = 0$, but can be larger than one for a range of values of $(\alpha, \theta)$, but that we have argued that regions with $\alpha<20^{\circ}$ are unlikely and contrived in section~\ref{sec:magorientation}.

In the absence of a detection, one can derive constraints on $g_{a\gamma\gamma}$ by comparing the measured value of $q_{\rm meas.}$ for a particular template, with the expected value $\braket{q}$ for that same template. The measured value is defined by 
\begin{equation}
q_{\rm meas.}(\theta, \alpha) = \frac{\textbf{F}(\theta, \alpha) \cdot \textbf{d}}{\sigma_N} .
\end{equation}
This must be compared to the expected value  $\braket{q(g_{\rm a \gamma \gamma},\theta, \alpha)}$ if a signal were present in the data 
\begin{equation}
\braket{q(g_{\rm a \gamma \gamma},\theta, \alpha)} = \frac{\sqrt{\textbf{S}(g_{a\gamma \gamma},\theta,\alpha) \cdot \textbf{S}(g_{a\gamma \gamma},\theta,\alpha) }}{\sigma_N}.
\end{equation}
In the perturbative limit of the conversion probability, which we have checked is always the case here, this is $\propto g_{a\gamma\gamma}$. Therefore, in order to impose a limit we calculate $\braket{q(g_{\alpha \gamma \gamma}^{\rm fid},\theta, \alpha)}$ for $g_{a\gamma\gamma}^{\rm fid}=10^{-10}\,{\rm GeV}^{-1}$ and exclude any value such that 
\begin{equation}
    g_{a\gamma\gamma}>g_{a\gamma\gamma}^{\rm fid}\sqrt{2q_{\rm meas.}(\theta,\alpha)\over \braket{q(g_{\alpha \gamma \gamma}^{\rm fid},\theta, \alpha)} }.
    \label{exclusion}
\end{equation}
The factor of two in (\ref{exclusion}) corresponds to the observed noise level being twice the equivalent signal for $g_{a\gamma\gamma}$ meaning that this will be a $2$ $(\approx 95\%$ confidence) upper limit on coupling constant. 

We have explained above that the typical values of $q_{\rm meas.}$ are higher than they should be due to the noise model not being perfect. This will mean that the upper limits we compute using (\ref{exclusion}) are not optimal and hence are slightly conservative. 

\begin{figure*}
    \centering
    \includegraphics[width=0.45\textwidth]{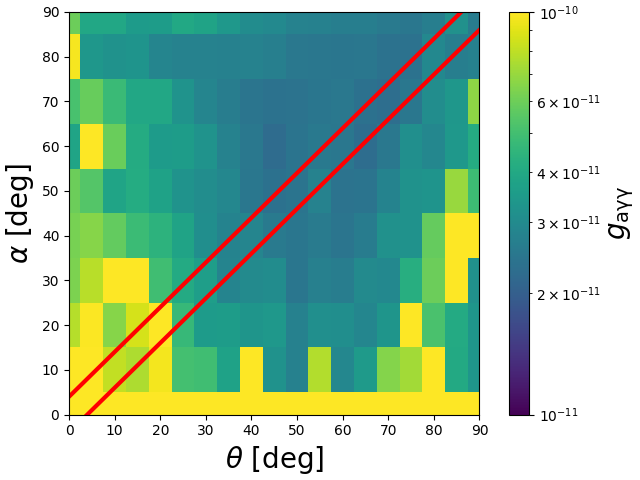}
    \includegraphics[width = 0.45\textwidth]{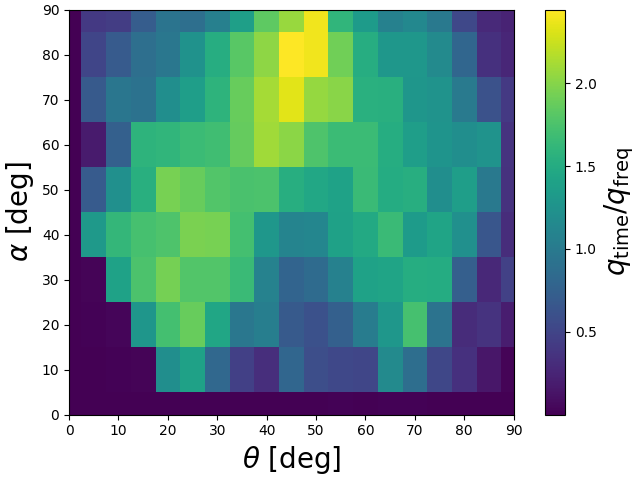}
    \caption{In the left panel, we present the constraints on the axion-photon coupling for the average-subtracted case as a function of $(\theta,\alpha)$ for $m_{\rm a}=4\,\mu{\rm eV}$. As expected, the derived limit is weaker in the limit $\alpha \rightarrow 0$ since the time-dependence of the signal is negligibly small. In fact there is no limit for $\alpha\equiv 0$ since there is no time dependence. Fortunately, the limits from the pulsar main beam modelling require $\alpha>20^{\circ}$ and $|\beta|<4^{\circ}$ which are included as red lines on the plot, so we are able to achieve a limit $g_{\gamma\gamma\gamma}\lesssim 9.6\times 10^{-11}\,{\rm GeV}^{-1}$. On the right panel, we show the ratio of the signal-to-noise in the case where the average has been subtracted (i.e., where only time-domain data is used) the case where the time-averaged flux is used in frequency space. In other words, this ratio quantifies the gain in working in the time-domain.}
    \label{fig:constraints}
\end{figure*}

We show the constraints from this procedure in the left panel of Fig.~\ref{fig:constraints} for $m_{\rm a}=4\,\mu{\rm eV}$ along with the regions preferred by our analysis of the orientation angles. In the right panel, we quantify the advantage of the purely time-domain analysis compared to the purely frequency-domain analysis as a function of $(\alpha, \theta)$. We do this by taking the ratio of theoretical signal-to-noise in the case where the baseline is subtracted, $q_{\rm time}$, with respect to the case where the time profile is averaged over the pulse-period and integrated, $q_{\rm freq.}$. When this ratio is larger than 1, the time-domain analysis leads to stronger constraints on $g_{a \gamma \gamma}$. 

\begin{figure}
    \centering
    \includegraphics[width = 0.45\textwidth]{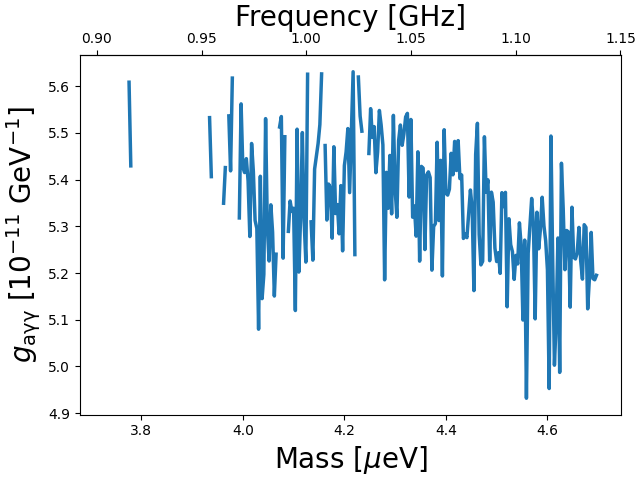}
    \caption{$2$ upper limits on $g_{a\gamma\gamma}$ as a function of $m_{\rm a}$ for $d=0.165\,{\rm pc}$. This is determined by calculating the highest upper limit in the region of the $\alpha-\theta$ plane allowed by $|\beta|<4^{\circ}$ and $\alpha>20^\circ$ from the equivalent of Fig. \ref{fig:constraints}. On the basis of this figure we quote an upper limit of $g_{a\gamma\gamma}<5.5\times 10^{-11}\,{\rm GeV}$ over the mass range $3.9\,\mu{\rm eV}\le m_{\rm a}\le 4.7\,\mu{\rm eV}$.}
    \label{fig:limit_plot}
\end{figure}

We have performed the same analysis as presented in Fig.~\ref{fig:constraints} over the mass range $3.9\,\mu {\rm eV}\le m_{\rm a}\le 4.7\,\mu{\rm eV}$ and then searched for the weakest upper limit in the range $\alpha>20^\circ$ and $|\beta|<4^\circ$. The results, assuming that $D=0.165\,{\rm kpc}$, are presented in Fig.~\ref{fig:limit_plot} and using this, we conclude that we can exclude dark matter axions forming the entire Galactic halo with $g_{a\gamma\gamma}>5.5\times 10^{-11}\,{\rm GeV}^{-1}$. If $D=0.29\, {\rm kpc}$ as suggested by electron density models, we find a weaker constraint of $g_{\rm a\gamma\gamma} > 9.6\times 10^{-11}\,{\rm GeV}^{-1}$. We leave the detailed investigation of the impact of the galactic electron density models and the veracity of the models for future work.

 \section{Future Searches}
\label{sec:future}

In the previous section we have shown how one can derive a limit on $g_{a\gamma\gamma}$ from baseline subtracted radio pulsar data using the variation in the time domain calculated by our ray-tracing algorithm. There we used a specific pulsar and telescope. An obvious question is what can be gained in future observations. In general, the limits obtained will scale as
\begin{equation}\label{eq:Radiometer}
g_{a\gamma\gamma}^{\rm lim}\propto \frac{1}{( \mu_S^2+ \sigma_S^2)^{1/4}}\left({A_{\rm eff}\over T_{\rm sys}}\right)^{-1/2}t_{\rm obs}^{-1/4}. 
\end{equation}
Hence, any improvement on $g_{a\gamma\gamma}^{\rm lim}$ will come from two avenues. The first is better observations: lower system temperature $T_{\rm sys}$, larger collecting area $A_{\rm eff}$ and increased observing time, $t_{\rm obs}$. The second is a better target: one with larger mean-signal power $\mu_S$ or greater time variability as measured by $\sigma_S$. Leveraging the latter is of course the key point of the present paper.  Let us now come to each of these factors in turn. 

In terms of improved observations with the current target, we can imagine future observations of PSR J2144$-$3933 with both MeerKAT or other similar telescopes in the short term and the Square Kilometre Array (SKA) in the longer term. All other things being equal, eq.~\eqref{eq:Radiometer} implies that an
increase in observation time from $\approx 1$ hour, as we have now, to 100 hours would improve the limit by a factor $\sim 3$. Moreover, going from MeerKAT with $A_{\rm eff}/T_{\rm sys}\approx 450$ to $A_{\rm eff}/T_{\rm sys}\approx 1800 $ for SKA1-mid will yield a limit which is a factor $\sim 2$ better. Combining these together
one might be able to obtain a limit of $g_{a\gamma\gamma}< 1.6\times 10^{-11}\,{\rm GeV}^{-1}$.

PSR J2144$-$3933 only allows constraints to be imposed for $m_{\rm a}<4.7\,\mu{\rm eV}$ within the framework of the GJ model. It might be interesting to perform lower frequency observations of it, but perhaps it is more interesting to find a source with a larger value of $m_{\rm a}^{\rm max}$ to probe mass ranges less accessible to terrestrial haloscopes. An interesting point to clarify would be how the signal scales as a function of pulsar parameters, in particular the period $P$ and the magnetic field $B_0$.

The figure of merit (FOM) based on radial trajectories is $\propto B_0^{2/3}P^{7/3}$, while the maximum mass probed is $\propto (B_0/P)^{1/2}$ \footnote{Note that this figure-of-merit is based on the calculation of power radiated at the poles for an aligned rotator Hook et al \cite{hook2018}. We think that this choice is a conservative estimate of the FOM. Indeed a time-varying FOM can be computed in the same approach, but we think the resultant ranking for targets from  such choices for the FOM are subject to errors arising from the radial trajectories assumption that has been superceded by the ray-tracing approach. }. At a fixed value of $P$ we see that it will always be best to increase the value of $B_0$, while at fixed $B_0$ the FOM will increase with $P$, but $m_{\rm a}^{\rm max}$ will decrease meaning that there is a trade-off between the two and the optimal target will be a compromise between the strength of the signal and the range of mass probed.

We have checked the scaling of the FOM using the ray-tracing code to calculate the average signal, $\mu_S$, summed over all frequencies for a range of values of $B_0$ and $P$. For $B_0$ these  appear to broadly confirm that the approximate scaling of the FOM  with relatively weak dependence $\theta$ and $\alpha$, although there can be significant deviations for extreme values. However, the dependence on $P$ seems to be somewhat weaker than that predicted from the FOM from radial trajectories. For the specific choice of $\alpha=60^{\circ}$ we find that $\hbox{FOM}\propto B_0^{0.8}P^{1.2}$. Given the complications in constructing a FOM which depends on the orientation angle, we conclude that it is reasonable to continue to use the FOM based on radial trajectories as a rule of thumb, but we should not expect it to give quantitatively accurate predictions for the increase in constraining power. The argument above applies to all attempts to constrain $g_{a\gamma\gamma}$ using neutron stars, not just a search for time dependent signals.

\begin{figure}
    \centering
    \includegraphics[width=0.45\textwidth]{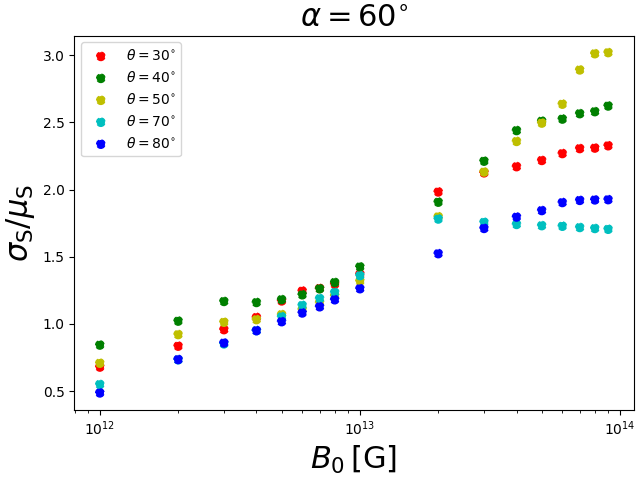}
    \caption{The relative time variance, $\sigma_S/\mu_S$, of the profiles as a function of the magnetic field $B_0$ at the surface of the neutron star. We fix $P=4\,{\rm s}$ and $m_{\rm a}= 1\,{\rm \mu eV}$ and the value of $g_{a\gamma\gamma}$ scales out in this ratio. Despite the time-variation between the maximum and minimum of the profiles increasing by orders of magnitude for $B_0 \sim 10^{14}\,{\rm G}$ compared to $B_0 \sim 10^{12}\,{\rm G}$, $\sigma_S/\mu_S$ only goes up by a factor of a few.}     
    \label{fig:sigma_mu}
\end{figure}

Let us come now to the question of other targets, focusing in particular on the time-dependence of their signals. Note that in our original target selection, we used only the mean power, however as demonstrated in Sec.~\ref{sec:mfmath} and as is apparent from Eq.~\eqref{eq:Radiometer}, the key parameter determining whether including a time-domain analysis can add value over just using the frequency domain is $\sigma_S/\mu_S$. Clearly this will depend on the intrinsic pulsar parameters, $B_0$, $P$ and axion mass, $m_a$ as well as $(\alpha,\theta)$. In Fig.~\ref{fig:sigma_mu} we show that $\sigma_S/\mu_S$ clearly increases with $B_0$ with the amount being sensitive to the choice of $\theta$ and $\alpha$.  We have also investigated the dependence on $P$, but this is typically much weaker for the relevant range of parameters. Based on this, we can expect that time-domain observations offer the greatest enhancement over a total flux measurement for larger values of $B_0$. Hence, objects such as magnetars stand to gain most from a time-domain versus total flux analysis. 

\begin{figure}
    \centering
    \includegraphics[width = 0.45\textwidth]{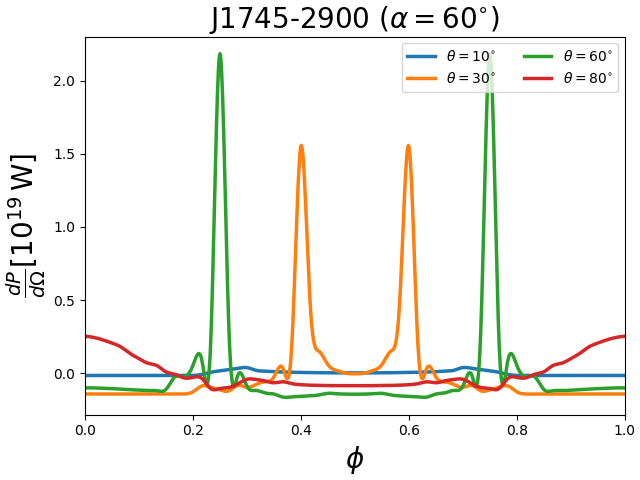}
    \caption{Predicted pulse profiles for the GCM. In order to generate these profiles, we fix $g_{a\gamma\gamma} = 10^{-10}\, {\rm GeV}^{-1}$ and $m_{\rm a} = 1\,\mu{\rm eV}$ and $\alpha=60^{\circ}$ while varying the observing angle. As in Fig.~\ref{fig:profiles} we have removed the average signal from the profiles. We can see very clearly that in some cases the profiles are very time variable - they almost look pulse-like for $\theta=30^{\circ}$ and $60^\circ$. Nonetheless these only correspond to a $\sigma_S/\mu_s\approx 3$ compatible with Fig.~\ref{fig:sigma_mu}.}
    \label{fig:magnetar_profiles}
\end{figure}

The GCM in particular has already been a popular target for searching for axion signals. In Fig.~\ref{fig:magnetar_profiles} we present some GCM profiles analogous to those in Fig.~\ref{fig:profiles}. The first thing to notice is that the signal is quite a bit larger, mainly due to the enhanced dark matter density assumed in the GC, although this is mitigated somewhat by it being further away. 

Clearly, at least for some choices of the orientation angles the profiles are substantially more localised in pulsar phase - they are almost pulse-like, but they are still much narrower than the width main peak in the radio pulse profile for this object. The strong dependence on pulse phase in this case is due to the effects of the ``throats'' in the magnetosphere where axion production is enhanced. This is likely to also enhance the constraining power. Obviously, this is a model-dependent assumption that arises from assuming the GJ model. We emphasise the point that the strength of this technique is directly proportional to the size of the time-dependence, and by extension the presence of these throats in the charge density.

Further to these points, the GCM offers both large relative time-variance and it lies in a region of larger dark matter density. Furthermore the axion-to-photon conversion is enhanced due to large magnetic fields. More specifically it has a large magnetic field $B_0\approx 1.4\times 10^{14}\,{\rm G}$ and a period of $P=3.76\,{\rm sec}$ and is a distance of $D\approx 8.3\,{\rm kpc}$. In fact, the ratio of the FOM Eq.~\eqref{eq:FOM} from PSR J2144$-$3933 relative to the GCM is given by 
\begin{equation}
{\hbox{FOM}|_{\rm GCM}\over\hbox{FOM}|_{J2144}}\approx 1.3\times 10^{-4}\times{\rho_{\rm DM}|_{\rm GCM}\over \rho_{\rm DM}|_{\rm local}}\,.
\end{equation} 
If one assumes a standard NFW profile for the galaxy, one obtains an enhancement of $\sim 10^5$ with respect to the local value \footnote{Note that without this assumption, the constraints from this pulsar are weaker than PSR J2144$-$3933.} which suggests that it will lead to a larger FOM by a factor of $\sim 20$. If the dependence on $B_0$ and $P$ is slightly stronger than using the radial trajectory based FOM, as we have suggested above,  we might expect a slightly stronger improvement than given by this simple argument. In addition, due to the larger value of $B_0$ from Fig.~\ref{fig:sigma_mu} we would expect $\sigma_S/\mu_S$ to be large enough to produce the most significant gains from a time-domain study compared to other targets. 

\begin{figure}
    \centering
    \includegraphics[width = 0.45\textwidth]{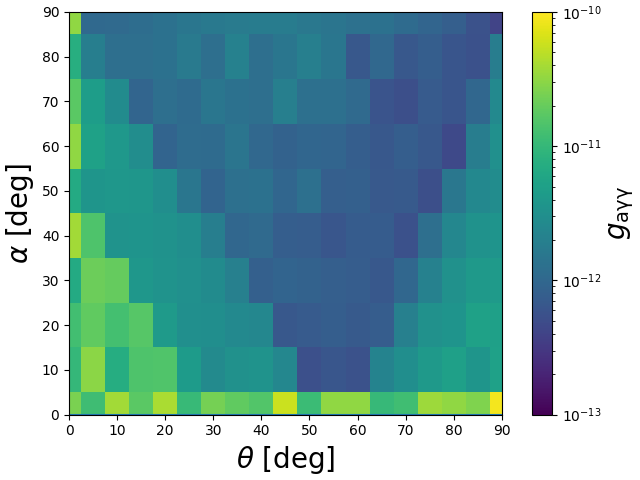}
    \includegraphics[width = 0.45\textwidth]{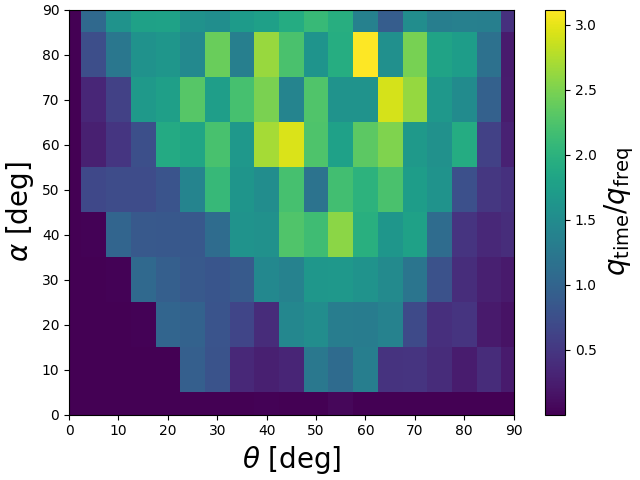}
    \caption{In the top panel, we show the \textit{expected} constraints on $g_{\rm a\gamma\gamma}$ from simulated observations of the GCM with an r.m.s. noise level of $3\,{\rm mJy}$ for $m_{\rm a} = 1\,\mu{\rm eV}$. We fix the pulsar parameters to be $B_0 = 1.4\times 10^{14}\,{\rm G}$, $P=3.76\, {\rm s}$ and $\rho_{\rm DM} = 5.4\times 10^{4}\,{\rm GeV\, cm^{-3}}$ which is the value computed using a standard NFW profile for the galaxy. Note that, depending on the values of $\theta$ and $\alpha$ limits as low as $g_{a\gamma\gamma}\lesssim 10^{-13},{\rm GeV}^{-1}$ might be possible. In the bottom panel, we quantify the effect of adding time-domain information over the parameter space $(\alpha,\theta)$. Note that this has a slightly different morphology to that for PSR J2144$-$3933, but also the values of $q_{\rm time}/q_{\rm freq}$ are slightly larger approaching $\approx 3$ as it is indicated might be the case in Fig.~\ref{fig:sigma_mu}.}
    \label{fig:magnetar}
\end{figure}

Prima-facie there seems to be an argument for reconsidering the GCM as a target. As an indication of what could be achieved for the GCM with similar observational resources to those currently available, we have simulated the equivalent of Fig.~\ref{fig:limit_plot} using the GCM as the source assuming a similar noise level to the present data, that is $3\,{\rm mJy}$, and this is presented in the top panel of Fig.~\ref{fig:magnetar}\footnote{We note that at 1.4\,GHz there is a significant increase in the sky temperature at the location of the GCM and this will dominate $T_{\rm sys}$. This means to achieve this noise figure one would need to observe for $\sim$10\,h with MeerKAT. In addition the pulsar and the axion signal will be scattered~\cite{Spitler:2014}. However, as the signal can be completely modelled using the observed pulsar profile it can be combined with the axion templates before the matched filtering is performed. Both these effects will fall off rapidly as a function of increasing frequency.}. Making similar assumptions about constraints on $(\alpha,\beta)$ from the pulse profile, we obtain a projected upper limit of $g_{a\gamma\gamma}<4\times 10^{-12}\,{\rm GeV}^{-1}$ which is a factor $\sim 10$ stronger than that we obtained from PSR J2144$-$3933. In addition, the GCM allows a much wider range masses with $m_{\rm a}^{\rm max}\approx 85\mu{\rm eV}$ $(f_{\rm obs}^{\rm max}\approx 20\,{\rm GHz})$. While there are many caveats to such an analysis, notably the noise levels that one might achieve in the direction of the GC, this suggests 
recording time domain information for this object  is well motivated. Note that above we have argued that it might be possible to improve limit by a factor $\sim 6$ using a 100 hour observation using SKA1-mid, which would lead to a projected limit of $g_{a\gamma\gamma}< 6\times 10^{-13}\,{\rm GeV}^{-1}$.

It is worth noting that the GJ model is believed to be a reasonable description of the pulsar magnetosphere for typical radio pulsars with magnetic field strengths in the range $10^{11}\,{\rm G} \leq B_0 \leq 10^{13}\,{\rm G}$. However, it breaks down in pulsars with large magnetic field values, magnetars in particular for which  the emission mechanism is poorly understood. Since a detailed analysis of the magnetosphere and the associated modelling uncertainties and propagating these uncertainties through our pipeline requires detailed MHD simulations, we leave such efforts to future work. It is a well-known fact that DM density at the galactic centre is very poorly constrained.  It is our considered opinion that the extrapolation of the NFW profile to the position of the GCM in the galactic halo to predict $\rho_{DM}$, which translates to an uncertainty of $\sim 2$ orders of magnitude in $g_{\rm a\gamma\gamma}$ is by far the biggest uncertainty in our GCM modelling. In order to address the former problem, we will now describe a model-independent analysis of the radio pulsar data to look for generalised periodic signals. 

In the bottom panel of Fig.~\ref{fig:magnetar}, similar to the right panel of Fig.~\ref{fig:constraints}, we show how much one can gain from doing the time-domain analysis as a function of $(\alpha, \theta)$. We see that the values of $q_{\rm time}/q_{\rm freq}$ larger, typically $\approx 2-3$ compared to $\approx 1$ for PSR J2144$-$3933. It seems reasonable to conclude that the factor $\approx 20$  improvement in the constraining power seen in the case of the GCM comes from the combination of the FOM based on radial trajectories, a slightly stronger dependence of the signal on $B_0$ and $P$ than predicted by radial trajectories and the use of the time domain structure. Our conclusion is that there is a strong argument for attempting to apply this technique to the GCM.

Finally, we comment that it goes without saying that improved knowledge of the orientation angles will lead to improved constraints. The constraints which we have imposed on $g_{a\gamma\gamma}$ from PSR J2144$-$3933 are strongly dependent on constraints we have placed on $\alpha$ and $\beta$ from the pulse profile. These are conservative and given the nature of the signal - there are some regions of the $\alpha-\theta$ plane where the constraints are very weak and even non-existent - meaning one is always dominated by the lower limit one imposes on $\alpha$. However, if one were to know the actual angles and, indeed, if they were in the region where the signal is predicted to be strongest then more optimal limits can be imposed. For example, in the case of the GCM if the actual angles are in the region where the strongest limit is, $g_{a\gamma\gamma}\lesssim 10^{-13}\,{\rm GeV}^{-1}$ for a 1 hour observation with MeerKAT. This could be a factor of 6 lower for 100 hours with the SKA.

\section{Search for generalized periodic signals}
\label{sec:periodic}

In the previous sections we have deduced limits on $g_{a\gamma\gamma}$ using PSR J2144-3933 and have discussed how this might be improved in the future using the GCM. The key qualitative feature of the predicted signal profiles is that their timescale is given by the pulse period $P$ - this is what allows us to use the pulsar data already folded at the pulse period.  One might be concerned that the precise predictions using the GJ model might be too simplistic given the complicated nature of the pulsar magnetosphere. However, taking the qualitative prediction that the signal is dominated by low harmonics of the period one might perform a generalised search for the periodic signals in the data. Of course, without specific connection to the physics, such a search cannot yield an upper limit on $g_{a\gamma\gamma}$, but does allow us to search more specifically for periodic would-be axion signals which might otherwise be missed due to modelling uncertainties. 

Any time-periodic data $d(t)$ can be written as a Fourier series
\begin{equation}
d(t) = \sum_{k= -\infty}^{\infty} a_k e^{2 \pi i  k t/P}\,,
\end{equation}
where $P$ is the signal period. The coefficients $a_k$ are then given by
\begin{equation}
    a_k = \frac{1}{P} \int _0^P dt  \, \, d(t) e^{ - 2 \pi i k t /P}\,.
\end{equation}
Reverting to the discrete case considered in this text, where the data is defined on $N$ discrete time-bins, with entries $d_q$, $q=0,..,N-1$,  the Fourier coefficients in the discrete limit can be written as
\begin{equation}
    a_k  = \frac{1}{N}\sum_{q=0}^{N-1}d_q \exp{\left(-\frac{2\pi i q k}{N}\right)}\,.
 \end{equation}
Note that in what follows we do not consider the $k=0$ mode associated to the time-average $\mu_S$ of the profiles which, in principle, has been already removed by the data processing. In addition, we comment that the parity properties of the profiles we predict for the axion signal imposes ${\rm Im}(a_k)=0$.

\begin{figure}
    \centering
    \includegraphics[width = 0.45\textwidth]{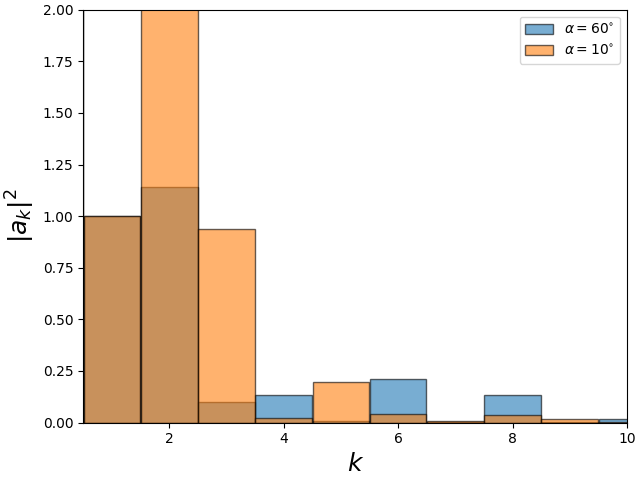}
    \medskip\medskip
    \includegraphics[width = 0.45\textwidth]{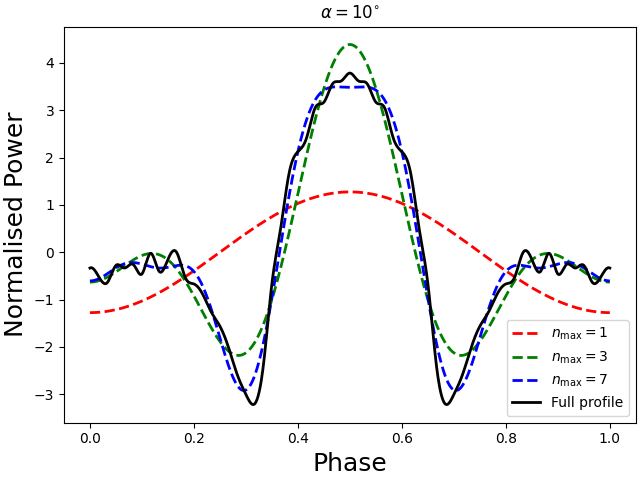}
    \medskip\medskip
    \includegraphics[width = 0.45\textwidth]{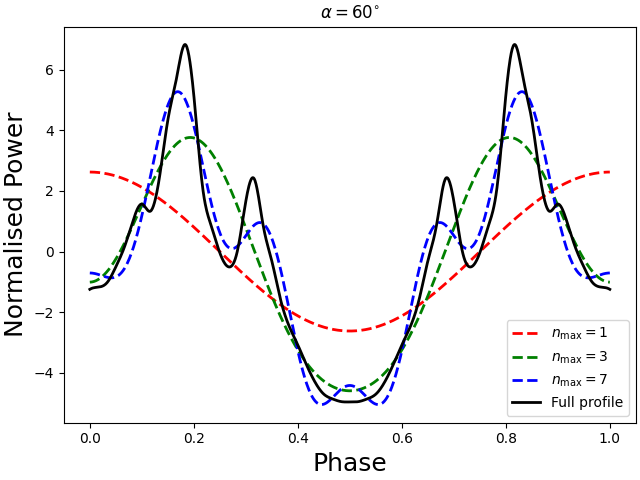}
    \caption{In the top panel we present the power spectrum of two profiles with with ($\alpha = 60^{\circ}, \theta = 30^{\circ})$ (blue) and $(\alpha = 10^{\circ}, \theta = 10^{\circ})$ (red). The power spectrum has been normalised such that $|a_1| = 1$. It is clear that both the power spectra are dominated by the lowest $n$ modes. In the middle and bottom panels we illustrate this point by presenting profiles obtained from the inverse transform of the FFT of each profile with $n_{\rm cut}=(1, 3, 7)$ (red, green and blue, respectively) compared to the exact profile (black). The significance of the low $n$ modes is further demonstrated by the fact that the blue curves are remarkably close in shape to the black curves. }
    \label{fig:FFT}
\end{figure}

In order to confirm our assumption that the profiles are dominated by low $k$ modes we have computed the power spectrum\footnote{We note that $a_k$ depends on the relative phase of the main peak of the pulsar profile and axion signal, but when one takes the power spectrum this information, which is encoded in the complex phase of $a_k$, is removed. So our test using the power spectrum is independent of this assumption.}, $|a_k|^2$, of two profiles with varying levels of time-dependence, i.e., a relatively flat profile with $(\alpha = 10^{\circ}, \theta = 10^{\circ})$ and one with a relatively large time variation with ($\alpha = 60^{\circ}, \theta = 30^{\circ})$. This is presented in the in the top panel of Fig.~\ref{fig:FFT}. We see that the power spectrum for both  decreases rapidly as a function of the mode number $k$. The point can be further reinforced by computing the inverse-Fourier transform from the $ a_{k}$ while neglecting all modes above some value of $k = n_{\rm cut}$. We have done this for $n_{\rm cut}=1,3,7$ and present the outcome in the middle and bottom panels of Fig.~\ref{fig:FFT}. It is clear that the profiles are reasonably well-represented by $n_{\rm cut}=3$ and $n_{\rm cut}=7$ in each case. We find that this is true for a wide range of predicted templates, but not all e.g. some of those presented in Fig.~\ref{fig:magnetar_profiles}. 

\begin{figure}
    \centering
    \includegraphics[width = 0.45\textwidth]{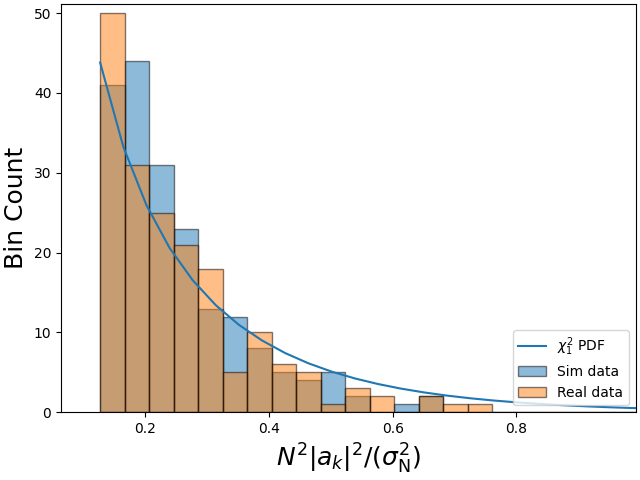}
    \includegraphics[width = 0.45\textwidth]{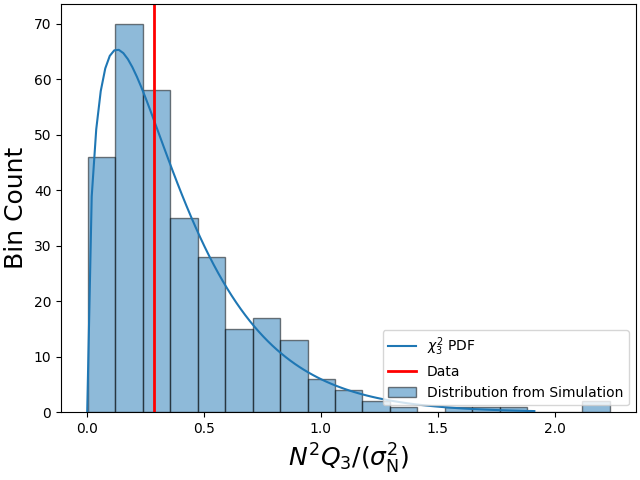}
    \includegraphics[width = 0.45\textwidth]{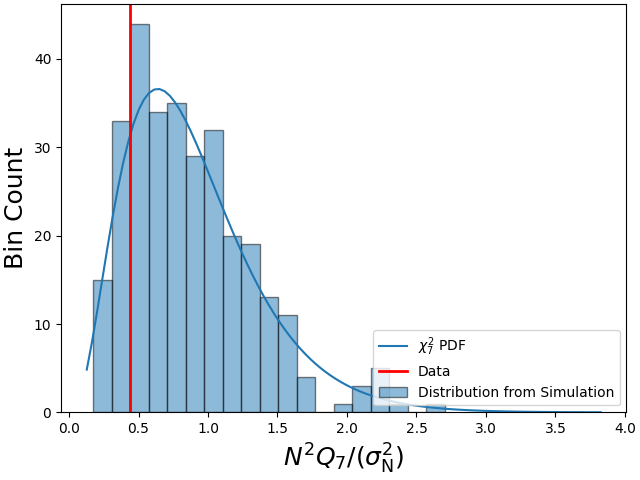}
    \caption{In the top panel we present a histogram of the power spectrum computed from the data for $m_{\rm a}=4\,\mu{\rm eV}$ in orange and a single Gaussian noise realization of the same variance and sample size. Each of the individual mode labelled by $k$ is binned, so this corresponds to a single mode search. We also plot the $\chi^2_1$ PDF associated expected for $|a_k|^2$ and observe that both the data, noise realization and analytic PDF are clearly compatible with each other. This appears to suggest that the data are compatible with being pure noise. In the middle and bottom panels we compare the distribution of the sum of first three modes (middle) and first seven modes (bottom) from 300 realisations of simulated Gaussian noise with the same standard deviation as the data. The measured value is represented with a red line. We use the same value $m_{\rm a}$ as in the top panel and have also included the theoretical PDFs which are $\chi^2_3$ and $\chi^2_7$ respectively. By doing this, we are searching for signals that have more structure than just a single mode in the time-domain. Since the data value is not a significant outlier, we exclude the presence of such signals in the data.}
    \label{fig:chisquared}
\end{figure}

The advantage of this approach is that we know a periodic signal search can be carried out using only the first $n_{\rm cut}$ Fourier modes, reducing computational overhead when scanning. 

Furthermore, in what follows, rather than scanning over every available template parametrised by $(a_1,\cdots, a_{n_{\rm cut}})$, with say, a fixed total power (which would be very costly from a numerical point of view) we instead compare the expected distribution of the $a_k$ that would follow if  $\textbf{d}$ were pure Gaussian noise. In that case the expected PDF for the values of the squared-amplitudes $|a_k|^2$ is a $\chi^2_m$ distribution 
\begin{equation}\label{eq:chisquared}
    {\cal P}(x;m) = \frac{1}{2^{m/2}\Gamma\left(\half m\right)}x^{m/2-1}\exp\left(-\half x\right)\, , 
\end{equation}
where $\Gamma$ is the standard gamma function. For a given mode, we have that $m=1$ is the number of degrees of freedom  since the noise is real, and hence the real and imaginary parts are correlated.  Note the distribution is the same for all $k$ meaning the spectrum is \textit{scale-invariant}. However, if there is an axion signal in the data, we expect this scale invariance to be broken, and, in particular,  for there to be a greater portion of power contained in the low Fourier modes, providing a model independent test of periodic axion signals in the data. Therefore, since the scale-invariance implies all $|a_k|^2$ have an identical distribution, by Fourier transforming the data $d(t)$ and binning the corresponding values of $|a_k|^2$ across all $k$, we can see to what extent they are $\chi_1^2$ distributed. This comparison is shown in the top panel of Fig.~\ref{fig:chisquared}, where we present histograms of the data (together with a single noise realization) for a particular frequency channel corresponding to a value of $m_{\rm a}=4\,\mu{\rm eV}$, both of which seem compatible with theoretical $\chi^2_1$ distribution. The sample is drawn from Fourier modes up to some $k_{\max}$ where the wavelength of the mode would be less than the temporal bin-width, beyond which the description breaks down owing to insufficient resolution. 

Having extracted spectral information from the data, we now want to analyse to what extent power is concentrated in the lower modes, as expected for an axion signal. To do this, we look at the power contained in the sum of the first $n_{\rm cut}$ modes, given by
\begin{equation}
Q(n_{\rm cut})=\sum_{k=1}^{n_{\rm cut}}|a_k|^2. 
\end{equation}
We then want to understand if the measured values of low-mode power are again consistent with the scale-invariant spectrum. The PDF for $Q(n_{\rm cut })$ is $\chi^2_{n_{\rm cut}}$ since it is the sum of $n_{\rm cut}$ independent modes each $\sim\chi^2_1 $.
This distribution can clearly be seen in the bottom two panels of Fig.~\ref{fig:chisquared} where for $n_{\rm cut}=3$ and $n_{\rm cut}=7$ we display simulated and analytic PDFs for the frequency channel corresponding to  $m_{\rm a}=4\,\mu{\rm eV}$. 
Note we have also included the measured value of $Q(n_{\rm cut})$ shown in red. 

In order to make a more precise statistical statement about the presence of a signal dominated by low-Fourier modes, we have calculated the probability, using the appropriate PDF, for there to be a larger value of $Q(n_{\rm cut})$ than that which is measured. This gives a sense of whether or not the measured value happens by chance in a way consistent with our sample size, or whether it is a sufficient outlier to indicate the presence of a periodic signal in the data. For the case of $n_{\rm cut}=3$ we find that there are three frequency channels where this ``probability to exceed'' $<0.05$ and one where it is $<0.02$.  However, since there are 213 individual frequency channels it seems likely that this is a chance outcome in a few frequency channels compatible with being a random noise realization. For $n_{\rm cut} = 7 $ there are none with a probability $<0.05$.

\section{Discussion and Conclusions}\label{sec:conclusions}

In this paper we have presented a procedure for carrying out time-domain searches for radio signals produced by axion dark matter converting into photons in the magnetospheres of pulsars. We have developed a matched filter formalism to define the signal-to-noise ratio of time-dependent signals and have used this to show that time-domain searches always improve the signal to noise ratio. By how much the SNR is improved is determined by the relative variance of the signal (the ratio of the standard deviation to the mean of the signal over the pulse period\footnote{Note that in pulsar astronomy, when applied to e.g. the main pulse, this quantity is referred to as the modulation index.}) and the matched filter formalism provides a robust framework to understand why this is the case. 

As a test case, we then applied the matched filter formalism to real data on PSR J2144$-$3933  obtained using MeerKAT, searching for expected periodic signal templates for the radio signatures produced by axion dark matter. This was selected from a list of pulsars on the basis of a simple figure of merit for axion detection. In the present analysis, for a fixed axion mass, these templates form a two-parameter family for each observing direction set by the angle $\theta$ between the stars rotation axis and the line of sight towards the pulsar, and $\alpha$ the angle between the stars magnetic axis and its rotation axis. Using the morphology of the observed pulsar main-beam signal, we were able to exclude a range of values $(\alpha,\theta)$, narrowing down the number of viable templates. Scanning over the allowed set of templates we find no evidence for axion dark matter and obtain an upper limit of $g_{a\gamma\gamma}< 9.6\times 10^{-11}\,{\rm GeV}^{-1}(D/0.29\,{\rm pc})$ over the mass range $3.9\,\mu{\rm eV}\le m_{\rm a}\le 4.7\,\mu{\rm eV}$. Given the astrophysical uncertainties in modelling the templates, we also carried out a generic periodic signal search independent of any modelling. This also returned no significant signal from axion dark matter.

\begin{figure*}[t!]
    \centering
    \includegraphics[width = 1.0  \textwidth]{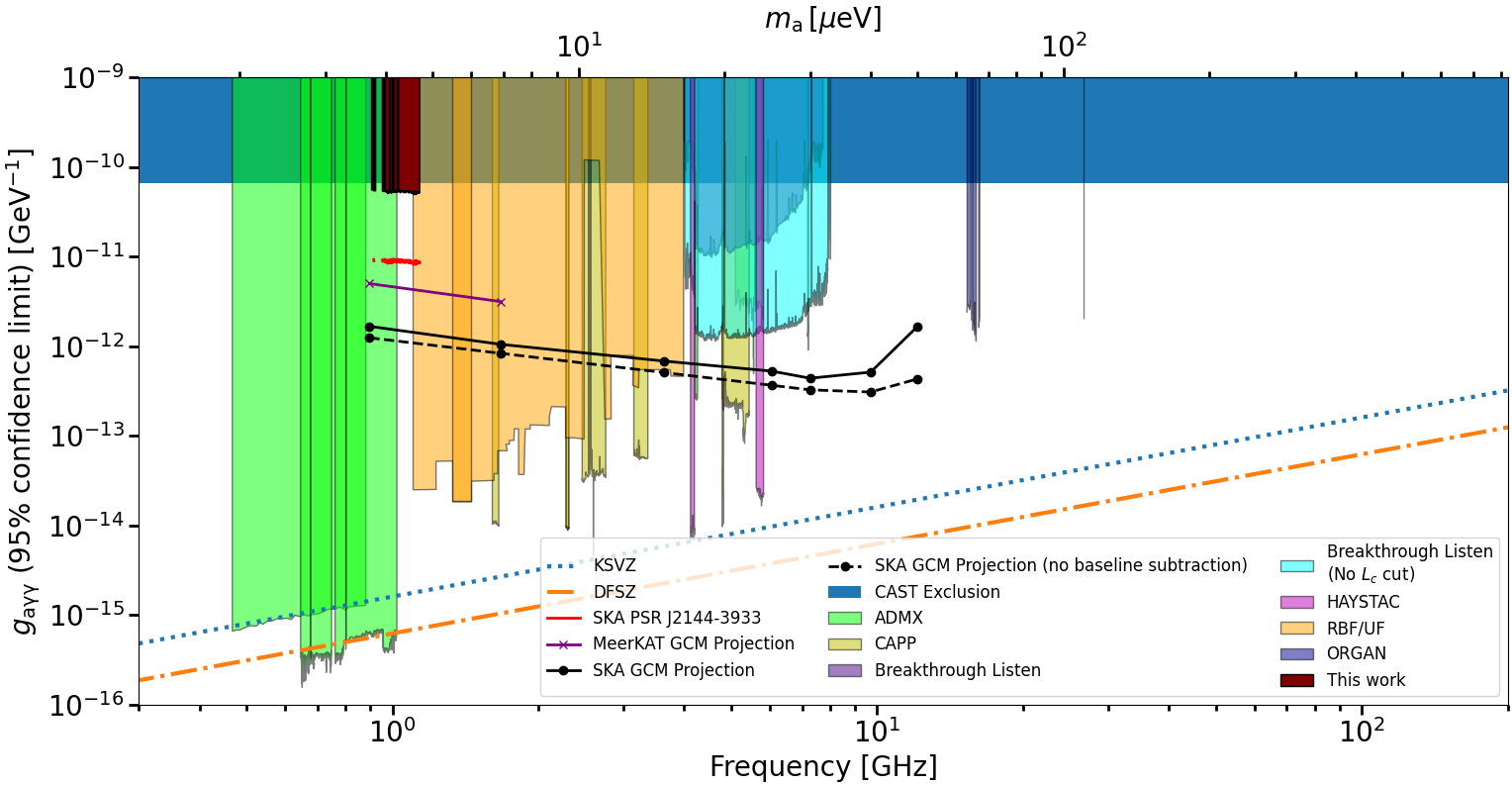}
    \caption{We show our limit (dark red) on the axion-photon coupling from 1 hour observations of PSR J2144-3933 ($B_0\simeq 2\times 10^{12}\,{\rm G}, \rho_{\rm DM} \simeq 0.45 \,{\rm GeV}{\rm cm}^{-3}$ and the projections from the 100 hours of observations from he SKA (red).  We also show projections of what can be achieved from $\sim 1$ hour observations of the GCM ($B_0 \simeq 1.6 \times 10^{14}\, {\rm G}$, $\rho_{\rm DM} \simeq 6.5 \times 10^{4}{\rm GeV}{\rm cm}^3)$ using  MeerKAT at L-band (purple) presented in this paper and 100 hours with the SKA (black) assuming a frequency coverage $0.9\leq f_{\rm obs} \leq 15.3$ GHz. Note that our projections for the GCM are derived assuming $\theta = \alpha = 20^{\circ}$ similar to what we assumed for PSR J2144-3933, which is conservative since the time-dependence can be larger for larger values of $\alpha$. Throughout we assumed an NFW profile. The dark blue exclusion region is from the CAST helioscope~\cite{ref:CAST}. The green and purple exclusions are from the ADMX~\cite{ADMX:2009iij,ref:ADMX2018,ADMX:2009iij,ADMX:2018gho,ADMX:2019uok,ADMX:2021nhd,ADMX:2018ogs,ADMX:2021mio,Crisosto:2019fcj} and HAYSTAC~\cite{HAYSTAC:Brubaker2017, HAYSTAC:2018rwy, HAYSTAC:2020kwv} haloscopes, respectively. The yellow exclusions are from the CAPP haloscope~\cite{CAPP:Jeong_2020cwz, CAPP:Lee_2020cfj, CAPP:Lee, CAPP:Lee_2022mnc, CAPP:Kim2022, Adair_2022} (Note that the high-frequency constraint  comes from the CAST-CAPP haloscope operating inside the CAST dipole magnet at CERN). The orange are from the RBF/UF haloscopes~\cite{RBF, UF}. Finally, the dark blue limits from the ORGAN collaboration~\cite{ORGAN1, ORGAN2}. We have also included the limits of the frequency only analysis of the pulsar population at the galactic centre using data from the Breakthrough Listen project~\cite{FosterSETI2022} as probably the most reliable limit, using all the known physics~\cite{Battye:2021xvt,Witte:2021arp} from neutron stars available prior to the analysis performed here. }
    \label{fig:exclusion_plot}
\end{figure*}
 
 In Fig.~\ref{fig:exclusion_plot} we have placed the limits derived here in context of other limits on $g_{a\gamma\gamma}$ from the CAST helioscope~\cite{ref:CAST}, haloscopes ~\cite{ADMX:2009iij,ref:ADMX2018,ADMX:2009iij,ADMX:2018gho,ADMX:2019uok,ADMX:2021nhd,ADMX:2018ogs,ADMX:2021mio,Crisosto:2019fcj,Brubaker2017, HAYSTAC:2018rwy, HAYSTAC:2020kwv,CAPP:Jeong_2020cwz, CAPP:Lee_2020cfj, CAPP:Lee, CAPP:Lee_2022mnc, CAPP:Kim2022,RBF, UF,ORGAN1, ORGAN2}. 
 There are a number of other limits in the literature from neutron stars~\cite{Foster:2020pgt,Darling:2020plz,darling2020apj,Battye_2020,FosterSETI2022,Zhou:2022yxp}. We have included~\cite{FosterSETI2022} as the present best limits obtained from observations that only use frequency information, but have not included the rest. For example, when it comes to GCM observations, the latest work by three of the present authors~\cite{Battye_2020}, was the most up-to-date since that work includes ray-tracing, however this was based on a previous version of ray tracing-code which, unlike the present work, did not include multiple reflections that increase the predicted signal, leading to overly-conservative constraints. The GCM will be re-analysed in \cite{McDonaldInPrep}, combining the most up-to-date account of the combined results on modelling from \cite{Witte:2021arp,Battye:2021xvt} and data~\cite{Darling:2020plz,darling2020apj,Battye2022}. Ref.~\cite{Zhou:2022yxp} also does not include ray-tracing so we do not include it. Similarly~\cite{Foster:2020pgt} (which we again do not include) has been superseded by the authors' follow-up work~\cite{FosterSETI2022} (shown in Fig.~\ref{fig:exclusion_plot}), which uses the most up-to-date modelling. We have tried to be fair in displaying those results which use the most up-to-date modelling and conservative assumptions. 

In this work, we set out with the intention of examining to what extent detailed time-domain information could be leveraged to increase the reach of radio searches for axions relative to  a simple radio-line search which simply averages the flux over a long-time. We were able to quantify this precisely in terms of the time-variance of the signal which we examined both for two specific pulsars and a range of other pulsar magnetic field strengths. It seems that for characteristic pulsar parameters there is a modest enhancement to the signal to noise from including time-domain information, however this marginal gain could be enough to tip a tentative detection in a total-flux measurement into a signal-to-noise level above 5 when time-domain information is included, making it well worthwhile to extract maximum leverage from pulsar observations. This is especially relevant as we look to future telescopes such as the SKA where we want to use all possible tools at our disposal to enhance the prospects for detection. Furthermore, given the rich variety of ever-increasing astrophysical probes of axions~\cite{Witte:2022cjj, Noordhuis:2022ljw,Chadha-Day:2022inf,Escudero:2023vgv,Day:2019bbh} events which are sharply peaked in time (or other detailed features amenable to a matched filter search) could benefit from more sophisticated search strategies. 

\begin{center}
\textbf{Acknowledgements}
\end{center}

JIM thanks Sam Witte for useful discussions and is supported by an FSR Fellowship. SS was supported formerly by a George Rigg Scholarship and more recently by the UK Science and Technology Facilities Council (STFC). Pulsar research at the Jodrell Bank Centre for Astrophysics is supported by a consolidated grant from
the STFC. The data used in this project was obtained through the MeerTIME project. We would like to thank that collaboration for making this data available to us.

\appendix

\section{Interpolation of the ray-tracing results}

\begin{figure}
    \centering
\includegraphics[width = 0.45\textwidth]{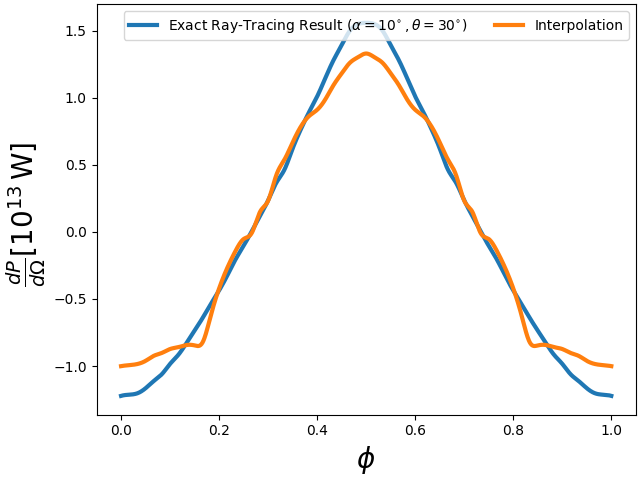}
    \includegraphics[width = 0.45\textwidth]{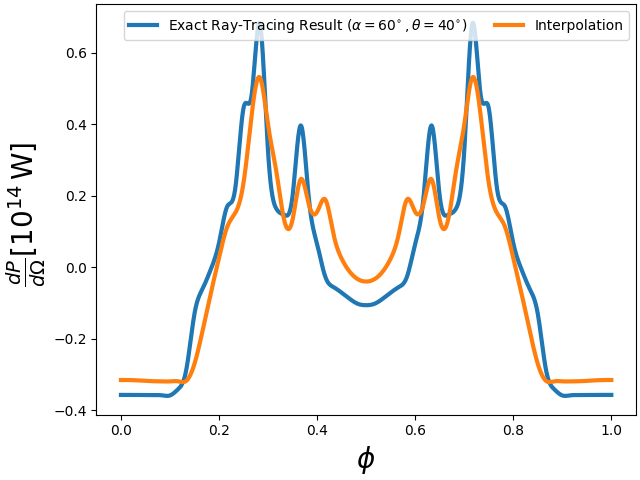}
    \includegraphics[width = 0.45\textwidth]{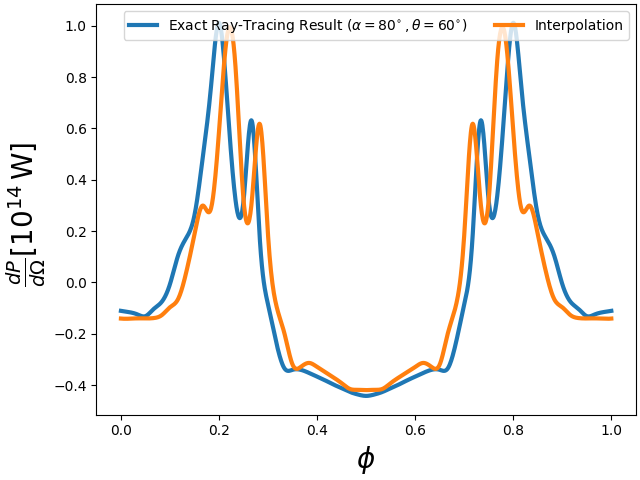}
    \caption{Examples of the interpolated profiles in the analysis. Each of these is for PSR J2144$-$3933 using the parameters discussed in the main text. The top panel is for $(\alpha,\theta)=(10^{\circ},30^{\circ})$, the middle for $(\alpha,\theta)=(60^{\circ},40^{\circ})$ and the bottom for $(\alpha,\theta)=(80^{\circ},60^{\circ})$. It is clear that the interpolation is not perfect, but it is accurate to within the level required for our analysis.}
    \label{fig:testinterp}
\end{figure}

 Due to the computational cost of the ray-tracing simulations, that require $\sim 24$ hours to produce a pulse-profile when parallelized over 32 CPU cores, we require a faster alternative to predict the time-dependence of the signal for arbitrary input angles. Therefore, we generate a simulated database of flux profiles as a discrete function of $(\theta, \alpha)$, represented by the circular data points in the top panels of Fig.~\ref{fig:testinterp} with $\Delta\theta=10^{\circ}$ and $\Delta\alpha=5^{\circ}$. Based on these datasets we generate an interpolation routine (where we use the \texttt{SciPy} package \texttt{scipy.interpolate}) that can then predict the signal for arbitrary values of $\alpha$ and $\theta$, the performance of which can be seen in the bottom panels of Fig.~\ref{fig:testinterp}, where we compare the prediction of our interpolation routine with the data points. For the purpose of our analysis, the level of agreement is sufficient.

\bibliographystyle{apsrev4-1}
\bibliography{ref.bib}

\end{document}